\documentclass[a4paper,10pt]{article}
\pdfoutput=1

\usepackage[utf8]{inputenc}

\usepackage{ifpdf}
\usepackage{amsmath}
\usepackage{amsthm}
\usepackage{amsfonts}
\usepackage{graphicx}

\ifpdf
    \usepackage[pdftex]{hyperref}
\else
    \usepackage[dvips]{hyperref}
\fi

\newcommand\auxfun[1]{\expandafter\newcommand\csname #1\endcsname{%
 \mathop{\hbox{\rm #1}}\nolimits}}
\newcommand\af[1]{\mathop{\hbox{\rm \textsc{#1}}}\nolimits}

\newcommand\meet{\sqcap}

\newcommand\op{\diamond}

\newtheoremstyle{break}%
{\topsep}{\topsep}
{\normalfont}
{}
{\scshape}
{:}
{\newline}
{}

\newtheoremstyle{property}%
{\topsep}{\topsep}
{\normalfont}
{}
{\slshape}
{:}
{ }
{}

\theoremstyle{plain}	 
\theoremstyle{break}  
\theoremstyle{remark}	
\theoremstyle{property} 

\numberwithin{Prop}{thm}

\newtheorem{definition}{Definition}

\title{A Survey of Distributed Data Aggregation Algorithms}

\author{Paulo Jesus, Carlos Baquero, Paulo Sérgio Almeida
\\\\\small{University of Minho (CCTC-DI)}\\\small{Campus de Gualtar, 4710-057 Braga, Portugal}\\\texttt{\small\{pcoj, cbm, psa\}@di.uminho.pt}}
\date{Technical Report\\September, 2011}

\begin{document}
	
\maketitle

\begin{abstract}
	Distributed data aggregation is an important task, allowing the decentralized determination of meaningful global properties, that can then be used to direct the execution of other applications. The resulting values result from the distributed computation of functions like \textsc{count}, \textsc{sum} and \textsc{average}. Some application examples can found to determine the network size, total storage capacity, average load, majorities and many others. 
	In the last decade, many different approaches have been proposed, with different trade-offs in terms of accuracy, reliability, message and time complexity. Due to the considerable amount and variety of aggregation algorithms, it can be difficult and time consuming to determine which techniques will be more appropriate to use in specific settings, justifying the existence of a survey to aid in this task.
	This work reviews the state of the art on distributed data aggregation algorithms, providing three main contributions. First, it formally defines the concept of aggregation, characterizing the different types of aggregation functions. Second, it succinctly describes the main aggregation techniques, organizing them in a taxonomy. Finally, it provides some guidelines toward the selection and use of the most relevant techniques, summarizing their principal characteristics. 
\end{abstract}


%

%
%


\section{Introduction}

Data aggregation is an essential building block of modern distributed systems, enabling the determination of important system wide properties in a decentralized manner. The knowledge of these global properties can then be used as input by other distributed application and algorithms. The network size is a common example of such global properties, which is required by many algorithms in the context of Peer-to-Peer (P2P) networks, for instance: in the construction and maintenance of Distributed Hash Tables (DHT) \cite{Manku:2003p3892,Stoica:2001p3981}; to set the number of targets of a gossip protocol \cite{Ganesh:2003p4101}. An estimation of the system size is also used in many other contexts, for example: to set up a quorum in dynamic settings~\cite{Abraham:2005p5920}, or by a membership service for wireless ad hoc networks (more precisely, to compute the mixing time of a random walk) \cite{BarYossef:2008p8282}. The network size is computed through the \textsc{count} aggregation function. Nevertheless, other meaningful global properties can be computed using different functions, for example: \textsc{average} can be applied to determine the average system load which can be used to direct local load balancing decisions; \textsc{sum} allows the determination of totals values such as the total free disk space available in a file-sharing system. In the particular case of Wireless Sensor Networks (WSN), data gathering is only practicable if data aggregation is performed, due to the strict energy constraints found on such environments. 

The above examples intend to illustrate some of the main reasons that have motivated the research and development of distributed data aggregation approaches along the past years, but more can be found in the literature. Besides all the existing relevant application examples, aggregation has also being stated as one of basis for scalability in large scale services \cite{Renesse:2003p677}, reinforcing its importance. Currently, a huge amount of distinct approaches constitute the body of related work on distributed data aggregation algorithms, with all exhibiting different trade-offs in terms of accuracy, time, communication and fault-tolerance. All existing techniques have confirmed that obtaining global statistics in a distributed fashion is a difficult problem, specially when considering faults and network dynamism. Moreover, in front of such diversity, it becomes difficult to choose which distributed data aggregation algorithm should be preferred in a given scenario, and which one will best suit the requirements of a specific application. One of the main motivations of this work is precisely to help readers make this choice. 

Some surveys have been previously published~\cite{Fasolo:2007p2527,Rajagopalan:2005p7425,YingpengSang:2006p2526,Alzaid:2008p5615,Nakamura:2007p2536} focusing specifically on aggregation techniques for WSN. Several in-network aggregation techniques for WSN are depicted in~\cite{Fasolo:2007p2527}, which typically operate at the network level, needing to deal with the resource constraints of sensor nodes (limited computational power, storage and energy resources). A review of the existing literature more focused on energy efficiency is presented in~\cite{Rajagopalan:2005p7425}, and on security in~\cite{YingpengSang:2006p2526,Alzaid:2008p5615}. Another work reviewing the state-of-the-art of information fusion techniques for WSN is also available~\cite{Nakamura:2007p2536}. In this later work, a broader view of the sensor fusion process is reviewed, from raw data collection, passing through a possible summarization (data aggregation) or compression, until the final resulting decision or action is reached. Data aggregation is considered a subset of information fusion, that aims at reducing (summarize) the handled data volume.

In this survey, we intend to address data aggregation algorithms at a higher abstraction level, providing a comprehensive and more generic view of the problem independently of the type of network used. We define the problem of computing aggregation function in a distributed fashion, and detail a wide range of distinct solutions. Existing approaches are classified, and their advantages/disadvantage in terms of communication and computational complexity are discussed. Moreover, we give some important guidelines to help in the decision of which distributed aggregation algorithm should be use, according to the requirements of a target application and environment.

The remaining of this survey is organized as follows. In Section~\ref{sec:problem_definition}, we clarify the concept of aggregation, and define the problem of its distributed computation. A taxonomy of the existing distributed aggregation algorithms is proposed in Section~\ref{sec:taxonomy}, describing the most relevant approaches and discussing their \emph{pros} and \emph{cons}. Section~\ref{sec:practical_guidelines} summarize the properties of the most relevant approaches, and gives some guidelines for their practical application. Finally, some concluding remarks and future research directions are drawn in Section~\ref{sec:final_remarks_and_future_directions}.


\section{Problem Definition}
\label{sec:problem_definition}

In a nutshell, aggregation can be simply defined as ``the ability to summarize information'', quoting Robbert Van Renesse 
~\cite{Renesse:2003p677}. Data aggregation is considered a subset of information fusion, aiming at reducing (summarize) the handled data volume~\cite{Nakamura:2007p2536}. Here, we provide a more precise definition, and consider that the process consists in the computation of an \emph{aggregation function} defined by: 

\begin{definition}[Aggregation function]
An aggregation function $f$ takes a multiset of elements from a domain $I$ and produces an output of a domain $O$:
\[
f: \mathbb{N}^I \to O
\]
\end{definition}

The input being a multiset emphasizes that: 1) the order in which the elements are aggregated is irrelevant; 2) a given value may occur several times. Frequently, for common aggregation functions such as \textsc{min}, \textsc{max}, and \textsc{sum}, both $I$ and $O$ are the same domain. For others, such as \textsc{count} (which gives the cardinality of the multiset), the result is a nonnegative integer, regardless of the input domain.

An aggregation function aims to summarize information. Therefore, the result of an aggregation (in the output domain $O$) typically takes much less space than the multiset to be aggregated (an element from $\mathbb{N}^I$). We will leave unspecified what is acceptable for some function to be considered as summarizing information, and therefore, an aggregation function. It can be said that the output domain $O$ is not normally a multiset (we do not have normally $O = \mathbb{N}^I$) and that the identity function is clearly not an aggregation function as it definitely does not summarize information. In most practical cases, the size of the output is at most a logarithmic of the input size, and often even of constant size.

\subsection{Decomposable functions}

For some aggregation functions we may need to perform a single computation involving all elements in the multiset. For many cases, however, one needs to avoid such centralized computation. In order to perform distributed \emph{in-network} aggregation, it is relevant whether the aggregation function can be \emph{decomposed} into several computations involving sub-multisets of the multiset to be aggregated. For distributed aggregation it is useful, therefore, to define the notion of \emph{decomposable aggregation function}, and a subset of these, which we call \emph{self-decomposable aggregation functions}.

\begin{definition}[Self-decomposable aggregation function]
An aggregation function $f: \mathbb{N}^I \to O$ is said to be self-decomposable if, for some (merge) operator $\op$ and all non-empty multisets $X$ and $Y$:
\[
f(X \uplus Y) = f(X) \op f(Y)
\]
\end{definition}

In the above, $\uplus$ denotes the standard multiset sum (see, e.g.,\cite{Syropoulos:2001p9577}). According to the above definition, and given that the aggregation result is the same for all possible partitions of a multiset into sub-multisets, it means that the operator $\op$ is commutative and associative. Many traditional functions such as \textsc{min}, \textsc{max}, \textsc{sum} and \textsc{count} are self-decomposable:
\begin{eqnarray*}
\af{sum }(\{ x \}) & = & x, \\
\af{sum} ( X \uplus Y ) & = & \af{sum}(X) + \af{sum}(Y).
\end{eqnarray*}
\begin{eqnarray*}
\af{count}(\{ x \}) & = & 1, \\
\af{count}( X \uplus Y ) & = & \af{count}(X) + \af{count}(Y).
\end{eqnarray*}
\begin{eqnarray*}
\af{min} (\{ x \}) & = & x, \\
\af{min} ( X \uplus Y ) & = & \af{min}(X) \meet \af{min}(Y).
\end{eqnarray*}

\begin{definition}[Decomposable aggregation function]
An aggregation function $f: \mathbb{N}^I \to O$ is said to be
decomposable if for some function $g$ and a self-decomposable aggregation
function $h$, it can be expressed as:
\[
f = g \circ h
\]
\end{definition}

From this definition, the self-decomposable functions are a subset of the
decomposable functions, where $g = \af{id}$, the identity function.
While for self-decomposable functions the \emph{intermediate} results (e.g.,
for in-network aggregation) are computed in the output domain $O$, for a
general decomposable function, we may need a different auxiliary domain to
hold the intermediate results.

The classic example of a decomposable (but not self-decomposable) function,
is \textsc{average}, which gives the average of the elements in the multiset:
\begin{eqnarray*}
	\af{average}(X) &=& g(h(X)), \qquad \textrm{with} \\
h (\{ x \}) & = & (x,1) \\
h ( X \uplus Y ) & = & h(X) + h(Y), \\
g( (s,c) ) &=& s / c,
\end{eqnarray*}
in which $h$ is a self-decomposable aggregation function that outputs values
of an auxiliary domain (pairs of values)
and $+$ is the standard pointwise sum of pairs (i.e. $(x_1, y_1)+(x_2, y_2)=(x_1+x_2, y_1+y_2)$).
Another example is the \textsc{range} function in statistics, which gives the
difference between the maximum and the minimum value.

\subsection{Duplicate sensitiveness and idempotence}

Depending on the aggregation function, it may be relevant whether a given
value occurs several times in the multiset. For some aggregation functions,
such as \textsc{min} and \textsc{max}, the presence of duplicate values
in the multiset does not influence the result, which only depends on its
\emph{support set} (the set obtained by removing all duplicates from the
original multiset). E.g.,

\[
\af{min}(\{1, 3, 1, 2, 4, 5, 4, 5\}) = \af{min}(\{1, 3, 2, 4, 5 \}) = 1
\]

For others, such as \textsc{sum} and \textsc{count}, the number of times each
element occurs (its multiplicity) is relevant:

\[
8 = \af{count}(\{1, 3, 1, 2, 4, 5, 4, 5\}) \neq
\af{count}(\{1, 3, 2, 4, 5 \}) = 5
\]

Duplicate sensitiveness is relevant for distributed aggregation. Many
duplicate insensitive functions can be implemented using an idempotent binary 
operator on the elements of the multiset. This helps in obtaining fault
tolerance and decentralized processing, allowing retransmissions or
sending values across multiple paths.

\begin{definition}[Duplicate insensitive aggregation function]
An aggregation function $f$ is said to be duplicate insensitive if for all
multisets $M$, $f(M) = f(S)$, where $S$ is the support set of $M$.
\end{definition}

Moreover, some duplicate insensitive functions (like \textsc{min} and \textsc{max}) can be implemented using an idempotent binary operator, that can be successively applied (by intermediate nodes) on the elements of the multiset (any number of times without affecting the result). This helps in obtaining fault tolerance and decentralized processing, allowing retransmissions or sending values across multiple paths. Unfortunately, the distributed application of an idempotent operator is not always possible, even for some duplicate insensitive aggregation functions, such as \textsc{distinct count} (i.e. cardinality of the support set). In fact, the application of an idempotent operator in a distributed way to compute an aggregation function is only possible, if the function is duplicate insensitive and self-decomposable.

\subsection{Taxonomy of common aggregation functions}

Building on the concepts of decomposability and duplicate sensitiveness, we
can obtain a taxonomy of aggregation functions (see Table~\ref{tab:tax_agg_func}). This helps to clarify how
suited to distributed aggregation a function is. Non-decomposable functions
are harder than decomposable, and duplicate sensitive are harder than
duplicate insensitive. As we will see, one way to obtain fault-tolerance is
to use a duplicate insensitive approximation of some aggregation, that uses
an idempotent operation (like \textsc{max}) instead of a non-idempotent one (like
\textsc{sum}).

\begin{table}
\centering
\begin{tabular}{|c||c|c|c|}
	\hline
 & \multicolumn{2}{c|}{Decomposable} & Non-decomposable \\
	\hline
 & Self-decomposable & & \\
	\hline
	\hline
Duplicate insensitive & \textsc{min}, \textsc{max} & \textsc{range} & \textsc{distinct count}\\
	\hline
Duplicate sensitive & \textsc{sum}, \textsc{count} & \textsc{average} & \textsc{median}, \textsc{mode}\\
	\hline
\end{tabular}
\caption{Taxonomy of aggregation functions.}
\label{tab:tax_agg_func}
\end{table}

\section{Taxonomy} 
\label{sec:taxonomy}

In this section, a simple taxonomy of the existing distributed data aggregation algorithms is proposed, classifying them according to two main perspectives: communication and computation (see Table \ref{tab:tax_communication} and \ref{tab:tax_computation}). The first viewpoint refers to the routing protocols and intrinsic network topologies associated to each protocol, which are used to support the aggregation process. The second perspective points out the aggregation functions computed by the algorithms and the main principles from which they are based on. Other perspectives (e.g. algorithm requirements, covered types aggregation functions) could have been considered, since the mapping between the algorithms attributes is multidimensional. However, we believe that the two chosen perspectives will provide a clear presentation. 

\subsection{Communication}
Three major classes of aggregation algorithms are identified from the communication perspective, according to the characteristics of their communication pattern (routing protocol) and network topology (see Table \ref{tab:tax_communication}): \emph{structured} (usually, hierarchy-based), \emph{unstructured} (usually, gossip-based), and \emph{hybrid} (mixing the previous categories). 

The \emph{structured} communication class refers to aggregation algorithms that are dependent on a specific network topology and routing scheme to operate correctly. If the required routing topology is not available, then an additional preprocessing phase is needed in order to create it, before starting the execution of the algorithm. This dependency limits the use of these techniques in dynamic environments. For instance, in mobile networks these algorithms need to be able to continuously adapt their routing structure to follow network changes. Typically, algorithms are directly affected by problems from the used routing structure. For example, in tree-based communication structures a single point of failure (node/link) can compromise the delivery of data from all its subtrees, and consequently impair the applications supported by that structure. In practice, hierarchical communication structures (e.g. tree routing topology) are the most often used to perform data aggregation, especially in WSN. Alternative routing topologies are also considered, like the ring topology, although very few approaches rely on it. 

The \emph{unstructured} communication category covers aggregation algorithms that can operate independently from the network organization and structure, without establishing any predefined topology. In terms of communication, this kind of algorithms is essentially characterized by the used communication pattern: \emph{flooding/broadcast}, \emph{random walk} and \emph{gossip}. The \emph{flooding/broadcast} communication patterns is associated to the dissemination of data from one node to all the network or group of nodes -- ``one to all''. A \emph{random walk} consists in sequential message transmissions, from one node to another -- ``one to one''. The \emph{gossip} communication pattern refers to a well known communication protocol, based on the spreading of a rumor~\cite{Pittel:1987p1583} (or an epidemic disease), in which messages are sent successively from one node to a selected number of peers -- ``one to many''. In the recent years, several aggregation algorithms based on gossip communication have been proposed, in an attempt to take advantages of its simplicity, scalability and robustness. More details about these different communication patterns (see Table \ref{tab:tax_communication}), used to perform data aggregation, will be further described. 

The \emph{hybrid} class groups algorithms that mix the use of different routing strategies from the previous categories, with the objective to combine their virtues and reduce their weakness, in order to obtain an improved aggregation approach when compared to their progenitors used individually. 

\begin{table}[t]
\centering
\begin{tabular}{|c|c|l|}
	\cline{2-3}
	\multicolumn{1}{c|}{ } & \textbf{Routing} & \textbf{Algorithms} \\
	\hline 
	 &  & TAG~\cite{Madden:2002p2402}, DAG~\cite{Motegi:2005p1017}, \\
	 &  & I-LEAG~\cite{Birk:2006p3907},\\
	 & \emph{Hierarchy} & Sketches~\cite{Considine:2004p676},\\
	 \multicolumn{1}{|c|}{\bf Structured} & \emph{(tree, cluster, multipath)} & RIA-LC/DC~\cite{YaoChungFan:2008p5364,Fan:2010p9070}, \\
	 &  & Tributary-Delta~\cite{Manjhi:2005p7423}, \\
	 &  & Q-Digest~\cite{Shrivastava:2004p7854} \\
	 \cline{2-3}
	 \multicolumn{1}{|c|}{} & \emph{Ring} & (Horowitz and Malkhi, 2003)~\cite{Horowitz:2003p672}\\
     \hline
	 \hline
	 & \emph{Flooding/Broadcast} & Randomized Reports~\cite{Bawa:2003p673} \\
	\cline{2-3}
	 &  & Random Tour~\cite{Massoulie:2006p4521}, \\
	 & \emph{Random walk} & Sample \& Collide~\cite{Ganesh:2007p745,Massoulie:2006p4521},\\
	 &  & Capture-Recapture~\cite{Mane:2005p659}\\
	\cline{2-3}
	 &  & Push-Sum Protocol~\cite{Kempe:2003p700}, \\
	 \multicolumn{1}{|c|}{\bf Unstructured} &  & Push-Pull Gossiping~\cite{Jelasity:2005p664}, \\
	 &  & DRG~\cite{JenYeuChen:2006p3916},\\
	 &  & Flow Updating~\cite{Jesus:2009p8366,Jesus:2010p9117},\\
	 & \emph{Gossip} & Extrema Propagation~\cite{Baquero:2009p7919}\\
	 &  & Equi-Depth\cite{Haridasan:2008p8141}, \\
	 &  & Adam2~\cite{Sacha:2010p9280},\\
	 &  & Hop-Sampling~\cite{Kostoulas:2005p682,Kostoulas:2007p9773},\\
	 &  & Interval Density~\cite{Kostoulas:2005p682,Kostoulas:2007p9773}\\
	\hline
	\hline
	\multicolumn{1}{|c|}{\bf Hybrid} & \emph{Hierarchy $+$ Gossip} & (Chitnis et al., 2008)~\cite{Chitnis:2008p6744}\\
	\hline
\end{tabular}
\caption{Taxonomy from a communication perspective.}
\label{tab:tax_communication}
\end{table}

\subsubsection{Hierarchy-based approaches}
\label{sec:tree_based}

Traditionally, existing aggregation algorithms operate on a hierarchy-based communication scheme. Hierarchy-based approaches are often used to perform data aggregations, especially in WSN. This routing strategy consists on the definition of a hierarchical communication structure (e.g. spanning tree), rooted at a single point, commonly designated as \emph{sink}. In general, in a hierarchy-based approach the data is simply disseminated from level to level, up the hierarchy, in response to a query request made by the sink, which computes the final result. Besides the sink, other special nodes can be defined to compute intermediate aggregates, working as aggregation points that forward their results to upper level nodes until the sink is reached. Aggregation algorithms based on hierarchic communication usually work in two phases, involving the participation of all nodes in each one: \emph{request} phase and \emph{response} phase. The \emph{request} phase corresponds to the spreading of an aggregation request throughout all the network. Several considerations must be taken into account before starting this phase, depending on which node wants to performs the request and on the existing routing topology. For instance: if the routing structure has not been established yet, it must be created and ought to be rooted at the requesting node; if the required topology is already established, first the node must forward its request to the root, in order to be spread (from the sink) across all the network. During the \emph{response} phase, all the nodes answer the aggregation query by sending the requested data toward the sink. In this phase, nodes can be asked to simply forward the received data or to compute partial intermediate aggregates to be sent.

The aggregation structure of hierarchy-based approaches provides a simple strategy, that enables the exact computation of aggregates (without failures), in an efficient manner in terms of energy consumption. However, in adverse environments this type of approach exhibits some fragility in terms of robustness, since a single point of failure can jeopardize the obtained result. Furthermore, to correctly operate in dynamic environments, where the network continuously changes (nodes joining/leaving), extra resources are required to maintain an updated routing structure.

\paragraph{TAG}
The Tiny AGgregation service for ad-hoc sensor networks described by Madden et al.~\cite{Madden:2002p2402} represents a classical tree-based in-network aggregation approach. As referred by the authors, in a sense TAG is agnostic to the implementation of the tree-based routing protocol, as far as it satisfies two important requirements. First, it must be able to deliver query requests to all the network nodes. Second, it must provide at least one route from every node (that participates in the aggregation process) to the sink, guaranteeing that no duplicates are generated (at most one copy of every message must arrive). This algorithm requires the previous creation of a tree-based routing topology, and also the continuous maintenance of such routing structure in order to operate over mobile networks.

TAG supplies an aggregation service inspired in the selection and aggregation features of database query languages, providing a declarative SQL-like (Structured Query Language) query language to the users. This algorithm offers grouping capabilities and implements basic database aggregation functions, among others, such as: \textsc{count}, \textsc{maximum}, \textsc{minimum}, \textsc{sum} and \textsc{average}. The aggregation process consists of two phases: a \emph{distribution phase} (in which, the aggregation query is propagated along the tree routing topology, from the root to the leaves) and a \emph{collection phase} (where the values are aggregated from the children to the parents, until the root is reached). The obtention of the aggregation result at the root incurs a minimum time overhead that is proportional to the tree depth. This waiting time is needed to ensure the conclusion of the two execution phases and the participation of all nodes in the aggregation process. 

A \emph{pipelined aggregate} technique (detailed in~\cite{Madden:2002p695}) has been proposed to minimize the effect of the waiting time overhead. According to this technique, smaller time intervals (relatively to the overall needed time) are used to repetitively produce periodic (partial) aggregation results. In each time interval, all nodes that have received the aggregation request will transmit a partial result, which is calculated from the application of the aggregation function to their local reading and to the results received from their children in the previous interval. Along time, after each successive time interval, the aggregated value will result from the participation of a growing number of nodes, increasing the reliability and accuracy of the result, becoming close to the correct value at each step. The correct aggregation result should be reached after a minimum number of iterations (in an ideal fail-safe environment).

Following the authors concerns about power consumption, additional optimization techniques were proposed to the TAG basic approach, in order to reduce the number of messages sent, taking advantages of the shared communication medium in wireless networks (which enables message snooping and broadcast) and giving decision power to nodes. They proposed a technique called \emph{hypothesis testing}, to use in certain classes of aggregates, where each node can decide to transmit the value resulting from its subtree, only if it will contribute to the final result.

\paragraph{DAG}
An aggregation scheme for WSN based on the creation of a DAG (Directed Acyclic Graph) is proposed in ~\cite{Motegi:2005p1017}, with the objective to reduce the effect of message loss of common tree-based approaches by allowing nodes to possess alternative parents. The DAG is created by setting multiple parents (within radio range) to each node, as its next hop toward the sink. In more detail, request messages are extended with a list of parent nodes (IDs), enabling children to learn the parent's parent (grandparents) which are two hops away. In order to avoid duplicated aggregates, only a parent is chosen to aggregate intermediate values, preferably a common parent of its parents. The most common parent's parent between the list received from parents is chosen as the destination aggregator, otherwise one of the parents is chosen (e.g. when a node has only one parent node). Response messages are handled according to specific rules to avoid duplicate processing: they can be aggregated, forwarded or discarded. Messages are aggregated if the receiving node corresponds to the destination, forwarded if the destination is a node's parent, and discarded otherwise (destination is not the node or one of its parents). Note that, although the same message can be duplicated and multiple ``copies'' can reach the same node (a grandparent), they will have the same destination node and only one of them (from the same source) will be considered for aggregation (after receiving all messages from children). 

This method takes advantage of the path redundancy introduced by the use of multiple parents to improve the robustness of the aggregation scheme (tolerance to message loss), when compared to traditional tree-based techniques. Though a better accuracy can be achieved, it comes at the cost of an higher energy consumption, as more messages with an increased size are transmitted. Note that this approach does not fully overcome the message loss problem of tree routing topologies, as some nodes may have a single parent, being dependent from the quality of the created DAG.

\paragraph{Sketches}
An alternative multi-path based approach is proposed in~\cite{Considine:2004p676} to perform in-network aggregation for sensor databases, using small sketches. The defined scheme is able to deal with duplicated data upon multi-path routing and compute duplicate-sensitive aggregates, like \textsc{count}, \textsc{sum} and \textsc{average}. This algorithm is based on the probabilistic counting sketches technique introduced by Flajolet and Martin~\cite{Flajolet:1985p2833} (FM), used to estimate the number of distinct elements in a data collection. A generalization of this technique is proposed to be applied to duplicate-sensitive aggregation functions (non-idempotent), namely the \textsc{sum}. The authors consider the use of multi-path routing to support communication failures (links and nodes), providing several possible paths to reach a destination. Like common tree-based approaches, the algorithm consists of two phases: first, the sink propagates the aggregation request across the whole network; second, the local values are collected and aggregated along a multi-path structure from the children to the root. In this particular case, during the request propagation phase, all nodes compute their distance (level) to the root and store the level of their neighbors, establishing a hierarchical multi-path routing topology (similar to the creation of multiple routing trees). In the second phase, partial aggregates are computed across the routing structure, using the adapted counting sketch scheme, and sent to the upper levels in successive rounds. Each round corresponds to a hierarchy level, in which the received sketches from children nodes are combined with the local one, until the sink is reached. In the last round, the sink merges the sketches of its neighbors and produces the final result, applying an estimation function over the sketch. Notice that the use of an auxiliary structure to summarize all data values (FM sketches), and correspondent estimator, will introduce an approximation error that will be reflected in the final result. However, according to this aggregation scheme, it is expected that data losses (mitigated with the introduction of multiple alternative paths) will have an higher impact in the result accuracy than the approximation error introduced by the use of sketches (to handle duplicates).

\paragraph{I-LEAG}
This cluster-based aggregation approach, designated as I-LEAG~\cite{Birk:2006p3907} (Instance-Local Efficient Aggregation on Graphs), requires the pre-construction of a different routing structure -- \emph{Local Partition Hierarchy}, which can be viewed as a logical tree of local routing partitions. The routing structure is composed by a hierarchy of clusters (partitions), with upper level clusters comprising lower level ones. A single pivot is assigned to each cluster, and the root of the tree corresponds to the pivot of the highest level cluster (that includes all the network graph). This algorithm emphasizes local computation to perform aggregation, being executed along several sequential phases. Each phase, correspond to a level of the hierarchy, in which the algorithm is executed in parallel by all clusters of the corresponding level (from lower levels to upper levels). 

Basically, the algorithm proceeds as follow: each cluster checks for local conflicts (different aggregation outputs between neighbors); detected conflicts are reported to pivots, which compute the new aggregated value and multicast the result to the cluster; additionally, every node forwards the received result to all neighbors that do not belong to the cluster; received values are used to update the local aggregation value (if received from a node in the current cluster) or to update neighbor aggregation output (if received from a neighbor of the upper level cluster), enabling the local detection of further conflicts. 
Conflicts are only detected between neighbors that belong to a different clusters in the previous phase, with different aggregation outputs from those clusters. A timer is needed to ensure that all messages sent during some phase reach their destination by the end of the same phase. 

Further, two extension of the algorithm were proposed to continuously compute aggregates over a fixed network where node inputs may change along time: MultI-LEAG and DynI-LEAG~\cite{Birk:2006p2728}. MultI-LEAG mainly corresponds to consecutive executions of I-LEAG, improved to avoid sending messages when no input changes are detected. Inputs are sampled at regular time intervals and the result of the current sampling interval is produced before the next one starts. DynI-LEAG concurrently execute several instances of MultI-LEAG, pipelining its phases (ensuring that every partition level only executes a single MultI-LEAG phase at a time), and more frequently sampling inputs to faster track changes but at the cost of a higher message complexity. Despite the authors effort to efficiently perform aggregation, these algorithms are restricted to static networks (with fixed size), without considering the occurrence of faults.

\paragraph{Tributary-Delta}
This approach mixes the use of tree and multi-path routing schemes to perform data aggregation, combining the advantages of both to provide a better accuracy in the presence of communication failures~\cite{Manjhi:2005p7423}. Two different routing regions are defined: \emph{tributary} (tree routing, in analogy to the shape formed by rivers flowing into a main stem) and \emph{delta} (multi-path routing, in analogy to the landform of a river flowing into the sea). The idea is to use tributaries in regions with low message loss rates to take advantage of the energy-efficiency and accuracy of a traditional tree-based aggregation scheme, and use deltas in zones where message losses have a higher rate and impact (e.g., close to the sink where messages carry values corresponding to several node readings) to benefit from the multi-path redundancy of sketch based schemes. Two adaptation strategies (TD-Coarse and TD) are proposed to shrink or expand the delta region, according to the network conditions and a minimum percentage of contributing nodes predefined by the user. The prior knowledge of the network size is required, and the number of contributing nodes needs to be counted (or count the non contributing nodes in a tributary subtree), in order to estimate the current participation percentage. Conversion functions are also required to convert partial results from the tributary (tree-based aggregation) into valid inputs to be used in the delta region (by the multi-path algorithm). Experimental results applying TAG~\cite{Madden:2002p2402} in tributaries and Synopses Diffusion~\cite{Nath:2004p1114} (see Section \ref{sec:sketches}) in deltas, showed that this hybrid approach performs better when compared to both aggregation algorithms used separately.

\paragraph{Other approaches}
Several other hierarchy-based aggregation approaches can be found in the literature, most of them differing on the supporting routing structure, or on the way it is built. Beside alternative variations of the hierarchic routing topology, some optimization techniques to the aggregation process can also be found, especially to reduce the energy-consumption in WSN.

In \cite{Li:2005p685} an aggregation scheme over DHTs (Distributed Hash Tables) is proposed. This approach is characterized by its tree construction protocol, that use a \emph{parental function} to map a unique parent to each node, building an aggregation tree in a bottom-up fashion (unlike traditional approaches). The authors consider the coexistence of multiple trees to increase the robustness of the algorithm against faults, as well as the continuous execution of a tree maintenance protocol to handle the dynamic arrival and departure of nodes. Two operation modes are proposed to perform data aggregation (and data broadcast): \emph{default} and \emph{on-demand}. In the \emph{default} mode, the algorithm is executed in background, taking advantage of messages exchanged by the tree maintenance protocol (appending some additional information to these messages). The \emph{on-demand} mode corresponds to the traditional aggregation scheme found on tree-based algorithms. 

Zhao et al.~\cite{Zhao:2003p4475} proposed an approach to continuously compute aggregates in WSN, for monitoring purposes. They assume that the network continuously computes several aggregates, from which at least one corresponds to the min/max -- computed using a simple diffusion scheme. A tree is implicitly constructed during the diffusion process (node with the min/max value is set as the root of the created tree) and is used for the computation of other aggregates (e.g. average and count). In practice, two different schemes are used: a \emph{digest diffusion} algorithm to compute idempotent aggregates which is used to construct an aggregation tree, and a \emph{tree digest} scheme similar to common hierarchy-based approaches that operates over the tree routing structure created by the previous technique. 

Alternative hierarchic routing structures are found in the literature to support aggregation, namely: a BFS (Breadth First Search) tree is used in the GAP (Generic Aggregation Protocol)~\cite{Dam:2005p1714} protocol to continuously compute aggregates for network management purposes; the creation of a GIST (Group-Independent Spanning Tree) based on the geographic distribution of sensors is described in~\cite{Jia:2006p7372}, taking into consideration the variation of the group of sensors that may answer an aggregation query. A previous group-aware optimization technique has been proposed: GaNC (Group-Aware Network Configuration)~\cite{Sharaf:2004p2433}. GaNC influences the routing tree construction by enabling nodes to preferably set parents from the same group (analyzing the GROUP BY clause of the received aggregation queries) and according to a maximum communication range, in order to decrease message size and consequently reduce energy consumption. Some algorithms~\cite{Misra:2006p9123,Liao:2008p5112,Wang:2010p8893} based on swarm intelligence techniques, more precisely ant colony optimization, can also be found in the literature to construct optimal aggregation trees, once more to improve the energy efficiency of WSN. Ant colony optimization algorithms are inspired in the foraging behavior of ants, leaving pheromone trails that enable others to find the shortest path to food. In this kind of approach, the aggregation structure is iteratively constructed by artificial ant agents, consisting in the paths (from different sources to the sink) with the higher pheromone values, and where nodes nodes that belong to more than one path act as aggregation points.

Some studies~\cite{HongLuo:2006p8243,Heinzelman:2000p7944} have shown that deciding which node should act as a data aggregator or forwarder has an important impact on the energy-consumption and lifetime of WSN. 
A routing algorithm, designated AFST (Adaptive Fusion Steiner Tree), that adaptively decides which nodes should fuse (aggregate) data or simply forward it is described in~\cite{HongLuo:2006p8243}. AFST evaluates the cost of data fusion and transmission, during the construction of the routing structure in order to minimize energy consumption of data gathering. A further extension to this scheme was proposed to handle node arrival/departure, Online AFST~\cite{Luo:2009p5621}, with the objective of minimizing the cost and impact of dynamism in the routing structure. In LEACH (Low-Energy Adaptive Clustering Hierarchy)~\cite{Heinzelman:2000p7944,Heinzelman:2002p9398}, a cluster-based routing protocol for data gathering in WSN, the random rotation of cluster-heads along time is proposed in order to distribute the energy consumption burden of collecting and fusing (compressing) cluster's data.

Filtering strategies can also be applied to reduce energy consumption in hierarchy-based aggregation approach. For instance, A-GAP~\cite{Prieto:2006p2522} is an extension of GAP (previously referred) which uses filters to provide a controllable accuracy of the protocol. Local filters are added at each node in order to control whether or not an update is sent. Updates are discarded according to a predefined accuracy objective, resulting in a reduction in terms of communication overhead (number of messages). Filters can dynamically adjust along the execution of the protocol, allowing the control of the trade-off between accuracy and overhead. Another similar approach to reduce message transmissions according to a tolerated error value is proposed in ~\cite{Deligiannakis:2004p2341}, adaptively adjusting filters according to a Potential Gains Adjustment (PGA) strategy. A framework called TiNA (Temporal coherency-aware in-Network Aggregation) that filters reported sensor readings according to their temporal coherency was proposed in \cite{Sharaf:2004p2433}. This framework operates on the top of existing hierarchic-based aggregation schemes like TAG. In particular, TiNA defines an additional TOLERANCE clause to allow users to specify the desired temporal coherency tolerance of each aggregation query, and filter the reported sensor data (i.e. readings within the range of the specified value are suppressed).

\subsubsection{Ring based approaches}
\label{sec:ring}
Very few aggregation approaches are supported by a ring communication structure. This particular type of routing topology is typically surpassed by hierarchic ones, which are used instead. For instance, the effect of failures in a ring can be worst than on hierarchic topologies, as a single point of failure can break the all communication chain. Furthermore, the time complexity of rings to propagate data across all the network 
is typically higher, providing a slower data dissemination. However, this kind of topology can be explored in alternative ways, that can in some sense circumvent the aforementioned limitations.

It is worth referring to an alternative approach described by Horowitz and Malkhi~\cite{Horowitz:2003p672}, based on the creation of a virtual ring to obtain an estimation of the network size (i.e. \textsc{count}) at each node. This technique relies solely on the departure and arrival of nodes to estimate the network size, without requiring any additional communication. Each node of the network holds a single successor link, forming a virtual ring. It is assumed that each node possess an accurate estimator. Upon the arrival of a new node, a random successor among the existing nodes, named \emph{contact point}, is assigned to it. During the joining process, the new node gets the \emph{contact point} estimator and increments it (by one). At the end of the joining process, the two nodes (joining node and \emph{contact point}) will yield the new count estimate. Upon the detection of a departure, the inverse process is executed. This method provides a disperse estimative over the whole network, with an excepted accuracy that ranges from $n/2$ to $n^2$, where $n$ represent the correct network size. In the rest of this paper, we will always denote $n$ as the network size, unless explicitly indicated otherwise. Despite the achieved low accuracy and considerable result dispersion, this algorithm has a substantially low communication cost (i.e. communicates only upon arrival/departure, without any further information dissemination; each joining node communicates only with two nodes).

\subsubsection{Flooding/Broadcast based approaches}
\label{sec:flooding}
Flooding/Broadcast based approaches promote the participation of all network nodes in the data aggregation process. The information is propagated from a single node (usually a special one) to the whole network, sending messages to all neighbors -- ``one to all''. This communication pattern normally induces a high network load, during the aggregation process, implying in some cases a certain degree of centralization of data exchanges. Tree-based approaches are a traditional example of use of this communication pattern, but in this case supported by a hierarchic routing topology. Additional examples which are not sustained by any specific structured routing topology are described below.

\paragraph{Randomized Reports}
A naive algorithm to perform aggregation, will consist of broadcasting a request to the whole network (independently from the existing routing topology), collect the value at all nodes and compute the result at the starting node. This will likely lead to network congestion and an expected overload of the source node, due to \emph{feedback implosion}. However, a predefined response probability could be used to mitigate this drawback, such that network nodes will only decide to respond according to the defined probability. Such \emph{probabilistic polling} method was proposed in~\cite{Bawa:2003p673} to estimate the network size. The source node broadcast a query request with a sampling probability $p$, that will be used by all remaining nodes to decide whether to reply or not. All the received responses will be counted by the querying node (during a predefined time interval), knowing that it will receive a total number of replies $r$ according to the given probability. At the end, the network size $\widehat{n}$ can be estimated at the source by $\widehat{n} = r / p$.

\paragraph{Other Approaches}
A similar approach based on the same principle (sampling probability) is proposed in~\cite{Jurdzinski:2002p668}, to approximate the size of a single-hop radio network, considering the occurrence of collisions. In each step, a transmission succeeds if exactly a single station chooses to send a message. Setting a probability $p$ to decide to send a message at each node, the expression $np(1-p)^{n-1}$ gives the probability $p_s$ of a step being successful. The previous expression is maximized for $p=1/n$ where $p_s \approx 1/e$. It is expected that $p_s \approx t/e$, if the experiment is repeated independently $t$ times. Based on the previous probabilistic observation, the algorithm counts the number of successful steps along successive phases, to estimates the network size. Different probability values (decrementing) and number of trials (incrementing) are used along each consecutive phase, until the number of successful steps is close to the expected value. Further improvements to this algorithm have been proposed in~\cite{Kabarowski:2006p694}, aiming at making it immune against adversary attacks.

\subsubsection{Random Walk based approaches}
\label{sec:random_walk}
Random walk based approaches are usually associated to a data sampling process to further estimate an aggregation value, involving only a partial amount of network nodes. Basically, this communication process consists on the random circulation of a token. A message is sequentially sent from one node to another randomly selected neighbor -- ``one to one'', until a predefined stopping criteria is met (e.g. maximum number of hops, reach a selected node or return to the initial one). 
Usually, a small amount of messages are exchanged in this kind of approach, since only a portion of the network is involved in the aggregation process. Due to the partial participation of the network, algorithms using this communication pattern normally rely on probabilistic methods to produce an approximation of the computed aggregation function. Probabilistic methods provide estimations of the result with a known bounded error. If the execution conditions and the considered parameters of the algorithm are maintained, the estimation error is expected to be maintained (with constant bounds) along time. This kind of aggregation algorithms will not converge to the correct aggregation value, and the result will always contain an estimation error.

\paragraph{Random Tour}
The random tour approach~\cite{Massoulie:2006p4521} is based on the execution of a random walk to estimate a sum of functions of the network nodes, $\Phi = \sum_{i \in \mathcal{N}} \phi(i)$, for a generic function $\phi(i)$ where $i$ denotes a node and $\mathcal{N}$ the set of nodes (e.g. to estimate the network size, count: $\phi(i) = 1$, for all $i \in \mathcal{N}$). The estimate is computed from the accumulation of local statistics into a initial message, all of which are gathered during a random walk, from the originator node until the message returns to it. The initiator node $i$ initializes a variable $X$ with the value $\phi(i)/d_i$ (where $d_i$ denotes the degree of node $i$, i.e., number of adjacent nodes). Upon receive, each node $j$ adds to $X$ by $\phi(j)/d_j$ (i.e. $X \gets X + \phi(j)/d_j$). In each iteration, the message tagged with $X$ is updated and forwarded to a neighbor, chosen uniformly at random, until it returns to the initial node. When the originator receives back the message originally sent, it computes the estimate $\widehat{\Phi}$ (of the sum $\Phi$) by $\widehat{\Phi} = d_i X$.

\paragraph{Other approaches}
Other approaches based on random walks can be found in the literature, but they are commonly tailored for specific setting and to the computation of specific aggregation functions, like \textsc{count} (to estimate the network or group size). 

For instance, to accelerate self-stabilization in a group communication system for ad-hoc networks, a scheme to estimate the group size based on random walks is proposed in~\cite{Dolev:2006p3488} (first published in~\cite{Dolev:2002p3487}). In this specific case, a mobile agent (called \emph{scouter}) performs a random walk and collects information about alive nodes to further estimate the system size. The agent carries the set of all visited nodes and a counter associated to each one of them. Whenever the agent moves to a node, all the counters are incremented by one except the one of the current node, which is set to zero. Large counter values are associated to nodes that have been less recently visited by the \emph{scouter}, becoming more likely to be suspected of nonexistence. Counters are bounded by the scouter's maximum number of moves, which is set according to the expected cover time and a safety function, before considering a corresponding node as not connected. The main idea is to remove from the \emph{scouter} information of nodes -- sorted by increasing order of their counter value, where the gap between successive nodes ($k^{th}$ and $k-1^{th}$) is greater than the number of moves required to explore $k$ connected elements in a random walk fashion. After having the \emph{scouter} perform a large enough number of moves, the number of nodes in the system can be estimated by simply counting the number of elements kept in the set of visited nodes. 

Other relevant approaches based on the execution of random walks to collect samples, like \emph{Sample \& Collide}~\cite{Ganesh:2007p745,Massoulie:2006p4521} and \emph{Capture-Recapture}~\cite{Mane:2005p659}, are described in Section \ref{sec:counting}.

\subsubsection{Gossip-based approaches}
\label{sec:gossip_based}

Commonly, gossip and epidemic communication are indistinctly referred. However, in a relatively recent review of gossiping in distributed systems~\cite{Kermarrec:2007p3306} a slight distinction between the two is made. In a nutshell, the difference simply relies on the interaction directionality of both protocols. The authors state that gossiping is referred to ``the probabilistic exchange of information between two members'', and epidemic is referred to ``information dissemination where a node randomly chooses another member''. Even so, the effect of both protocols in terms of information spread is much alike, and strongly related to epidemics. Notice that, the information spread in a group in real life (gossip) is similar to the spread of an infectious disease (epidemics). For this reason, in this work no distinction will be made between gossip and epidemic protocols. 

Gossip communication protocols are strongly related to epidemics, where an initial node (``infected'') sends a message to a (random) subset of its neighbors (``contaminated''), which repeat this propagation process -- ``one to many''. With the right parameters, almost the whole network will end up participating in this propagation scheme. This communication pattern exhibits interesting characteristics despite its simplicity, allowing a robust (fault tolerant) and scalable information dissemination over all the network, in a completely decentralized fashion. Nevertheless, it is important to point out that the robustness of gossip protocols may not be directly attained by any algorithm based on a simple application of this communication pattern. For instance, an algorithm correctness may rely on principles and invariants that may not be guaranteed by a straightforward and incautious use of a gossip communication protocol, as revealed in~\cite{Jesus:2009p8115}. 
In general, gossip communication tends to be as efficient as flooding, in terms of speed and coverage, but it imposes a lower network traffic load (to disseminate data).

\paragraph{Push-Sum Protocol}
The push-sum protocol~\cite{Kempe:2003p700} is a simple gossip-based protocol to compute aggregation functions, such as \textsc{sum} or \textsc{average}, consisting of an iterative pairwise distribution of values throughout all the network. In more detail, along discrete times $t$, each node $i$ maintains and propagates information of a pair of values $(s_{ti}, w_{ti})$: $s_{ti}$ represents the sum of the exchanged values, and $w_{ti}$ denotes the weight associated to this sum at the given time $t$ and node $i$. In order to compute distinct aggregation functions, it is enough to assign appropriate initial values to these variables. 
E.g., considering $v_i$ as the initial input value at node $i$, \textsc{average}: $s_{0i} = v_i$ and $w_{0i} = 1$ for all nodes; \textsc{sum}:  $s_{0i} = v_i$ for all nodes, only one node starts with $w_{0i} = 1$ and the remaining assume $w_{0i} = 0$; \textsc{count}: $s_{0i} = 1$ for all nodes, only one with $w_{0i} = 1$ and the others with $w_{0i} = 0$. In each iteration, a neighbor is chosen uniformly at random, and half of the actual values are sent to the target node and the other half to the node itself. Upon receive, the local values are updated, adding each value from a received pair to its local component (i.e. pointwise sum of pairs). The estimate of the aggregation function can be computed by all nodes, at each time $t$ by $s_{ti}/w_{ti}$. The accuracy of the produced result will tend to increase progressively along each iteration, converging to the correct value. As referred by the authors, the correctness of this algorithm relies on a fundamental property defined as the \emph{mass conservation}, stating that: the global sum of all network values (local value of each node plus the value in messages in transit) must remain constant along time. Considering the crucial importance of this property, the authors assume the existence of a fault detection mechanism, that allow nodes to detect when a message did not reach its destination. In this situation, the ``mass'' is restored by sending the undelivered message to the node itself. This algorithm is further generalized by the authors in their work -- \emph{push-synopses protocol}, in order to combine it with random sampling to compute more ``complex'' aggregation functions (e.g. quantiles) in a distributed way. 

\paragraph{Other approaches}
In the last years, several gossip-based approaches have been proposed, due to the attractive characteristics of gossip communication: simplicity, scalability and robustness. Several alternative algorithms inspired by the \emph{push-sum protocol} have been proposed, like: \emph{Push-Pull Gossiping}~\cite{Jelasity:2004p662,Jelasity:2005p664} which provides an anti-entropy aggregation technique (see section \ref{sec:averaging}), or \emph{G-GAP}~\cite{Wuhib:2007p832} (Gossip-based Generic Aggregation Protocol) that extends the \emph{push-synopses protocol} to tolerate non contiguous faults (i.e. neighbors can not fail within the same short time period).

Another aggregation algorithm supported by an information dissemination and group membership management protocol, called \emph{newscast protocol}, is proposed in~\cite{Jelasity:2004p687}. This approach consists of the dissemination of a cache of items (with a predefined size) maintained by each network node. Periodically, each node randomly selects a peer, considering the network addresses of nodes available on the local cache entries. The cache entries are exchanged between the two nodes and the received information is merged into their local cache. The merge operation discards the oldest items, keeping a predefined number of the freshest ones, also ensuring that there is at most one item from each node in the cache. An estimate of the desired aggregate can be produced by each network node, by applying the aggregation function to the local cache of items.

\subsubsection{Hybrid approaches}
\label{sec:hybrid}

Hybrid approaches combine the use of different communication techniques to obtain improved results from their synergy. Commonly, the use of a hierarchic topology is mixed with gossip communication. Hierarchic based schemes are efficient and accurate, but highly affected by the occurrence of faults. On the other hand, gossip based algorithms are more resilient to faults, but less efficient in terms of overhead (requiring more message exchanges). In general, this combination enables hybrid approaches to achieve a fair trade-off between performance (in terms of overhead and accuracy) and robustness, when performing aggregation in more realistic environments.

\paragraph{(Chitnis et al., 2008)} 
Chitnis et al.~\cite{Chitnis:2008p6744} studied the problem of computing aggregates in large-scale sensor networks in the presence of faults, and analyzed the behavior of hierarchy-based (i.e. TAG) and gossip-based (i.e. Push-Sum Protocol) aggregation protocols. In particular, they observe that tree-based aggregation is very efficient for very small failures probabilities, but its performance drops rapidly with increasing failures. On the other hand, a gossip protocol is slightly slowed down (almost unaffected), and is better to use with failures (compared to tree-based). Considering these results, the authors proposed an hybrid protocol with the intent of leveraging the strengths of both analyzed mechanisms and minimize their weakness, in order to achieve a better performance in faulty large-scale sensor networks.

This hybrid approach divides the network nodes in groups, and a gossip-based aggregation is performed within each one. A leader is elected for each group, and an aggregation tree is constructed with the leader nodes (multi-hop routing may be required between leaders) to further perform a tree-based aggregation with the results from each gossip group. The authors also defined and solved an optimization problem to get the best combination between the two aggregation mechanisms, yielding the optimal size of the groups according to the network size and failure probability. However, in practice it requires the pre-computation of the gossip group size (by solving the referred optimization problem) before starting to use of the protocol with optimal settings. Results from simulations show that the hybrid aggregation approach usually outperforms the other two (tree-based and gossip-based) \footnote{Notice that only static network settings (no node arriving/leaving) were considered by the authors.}.

An extension of the previous approach for heterogeneous sensor networks is later discussed in~\cite{Chitnis:2009p6875}. In this case, it is considered that a few distinguished nodes, designated as \emph{microservers}, which are more reliable and less prone to failure than the remaining ones, are available in the network. The aggregation technique works mostly like the one previously described for the homogeneous case, but with two differences that take advantage of the reliability of \emph{microservers}. First, microservers are preferably chosen as group leaders. Second, microservers are put on the top of the created aggregation tree which may also be composed by other less reliable nodes. 
The use of microservers in the aggregation tree will increase its robustness, and by putting them at the top will reduce the need the reconstruct the whole tree when a fault occurs. The evaluation results show that the aggregation process can be enhanced in heterogenous networks, when taking advantage of more reliable (although more expensive) nodes.

\paragraph{Other Approaches} 	
A more elaborated structure was previously defined by Astrolabe~\cite{Renesse:2003p1272}. Astrolabe is a DNS-like distributed management system that supports attributes aggregation. It defines a hierarchy of zones (similar to the DNS domain hierarchy), each one holding a list of attributes called MIB (Management Information Base). This structure can be viewed as a tree, each level composed of non-overlapping zones, where leaf zones are single hosts, each one running an Astrolabe agent, and the root zone includes all the network. Each zone is uniquely identified by a name hierarchy (similarly to DNS), assigning to each zone a unique string name within the parent zone; the global unique name of each zone is obtained by concatenating the name of all its parent zones from the root with a predefined separator. The zone hierarchy is implicitly defined by the name administratively set to each agent. A gossip protocol is executed between a set of elected agents to maintain the existing zones. The MIB held by each zone is computed by a set of aggregation functions, that produce a summary of the attributes from the child zones. An aggregation function is defined by a SQL-like program that is code embedded in the MIB, being set as a special attribute. Agents keep a local copy of a subset of all MIBs, in particular of zones in the path to the root and siblings, providing replication of the aggregated information with weak consistency (eventual consistency). A gossip protocol is used for agents to exchange data about MIBs from other (sibling) zones and within its zone, and update its state with the most recent data. 

Another hierarchical gossiping algorithm was introduced by Gupta et al.~\cite{Gupta:2001p1478}, being one of the first to use gossip for the distributed computation of aggregation functions. According to the authors, the philosophy of this approach is similar to Astrolabe, but uses a more generic technique to construct the hierarchy, called \emph{Grid Box Hierarchy}. The hierarchy is created by assigning (random or topology aware) unique addresses to all members, generated from a known hash function. The most significant digits of the address are used to divide nodes into different groups (grid boxes) and define the hierarchy. Each level of the hierarchy corresponds to a set of grid boxes, matching a different number of significant digits. The aggregation process is carried out from the bottom to the top of the hierarchy in consecutive gossip phases (for each level of the hierarchy). In each phase: members of the same grid box gossip their data, compute the resulting aggregate after a predefined number of rounds, and then move to the next phase. The protocol terminates when nodes find themselves at the grid box at the top of the hierarchy (last phase). Note that, this approach does not rely on any leader election scheme to set group aggregators, in fact the authors argue the inadequacy of such mechanism in unreliable networks prone to message loss and node crashes.

Recently, an approach that combines a hierarchy based technique with random sampling was proposed in ~\cite{Cheng:2010p8605} to approximate aggregation functions in large WSN. In this approach, the amount of collect data is regulated by a sampling probability produced from the input accuracy (expressed by two parameters $\varepsilon$ and $\delta$, i.e., relative error less than $\varepsilon$ with probability greater than $1-\delta$) and the aggregation function (i.e. \textsc{count}, \textsc{sum} or \textsc{average}), aiming at reducing the energy consumption to compute the aggregate. This algorithm considers that the sensing nodes are organized in clusters (according to their geographic location), and that cluster heads form a spanning tree rooted at the sink. Basically, the aggregation proceeds as following: first, the sink computes the sampling probability $p$ (according to $\varepsilon$ and $\delta$) and transmits it along with the aggregation function to all cluster heads across the spanning tree; then, cluster heads broadcast $p$ to their cluster and each node within independently decides to respond according to the received probability; samples are collected at each cluster head which computes a partial result; finally, the partial results are aggregated upward the tree (convergecast) until the sink is reached, and where the final (approximated) result is computed. This algorithm, referred by the authors as BSC (Bernoulli Sampling on Clusters), mixes the application of a common hierarchy based aggregation technique such as \emph{TAG} (see Section~\ref{sec:tree_based})  between cluster heads, with a flooding/broadcast method like \emph{Randomized Reports} (see Section \ref{sec:flooding}) to sample the values at each cluster.

\subsection{Computation}
\label{sec:agg_techniques}

In terms of computational principles on which the existing aggregation algorithms are based, the following main categories (see Table \ref{tab:tax_computation}) were identified: \emph{Hierarchical}, \emph{Averaging}, \emph{Sketches (hash or min-k based)}, \emph{Digests}, \emph{Deterministic}, and \emph{Sampling}. These categories intrinsically support the computation of different kinds of aggregation functions. 
For instance, \emph{Hierarchical} approaches allow the computation of any decomposable function. \emph{Averaging} techniques allow the computation of all duplicate sensitive decomposable functions that can be derived from the \textsc{average}, by using specific initial input values and combining the results form different instances of the algorithms. \emph{Sketches} techniques also allow the computation of duplicate sensitive decomposable functions, but that can be derived from the \textsc{sum} \footnote{Note that, \textsc{count} is the sum of all elements considering their input value as equal to 1.}. Moreover, schemes based on hash sketches are natively able to compute distinct counts (non decomposable duplicate insensitive), and those based on min-k can be easily adapted to compute it (e.g. in extrema propagation, see \ref{sec:sketches}, using the input value as seed of the random generation function, so that duplicated values will generate the same number). 
\emph{Digests} support the computation of any kind of aggregation function, as this type of approach usually allows the estimation of the whole data distribution (i.e. values and frequencies) from which any function can be obtained. On the other hand, some techniques are restricted to the computation a single type of aggregation function, such as \textsc{count}, which is the case of \emph{Sampling} approaches.


\begin{table}[t]
\centering
\begin{tabular}{|c|c|c|l|}
	\hline
	\textbf{Aggregation} & \multicolumn{2}{|c|}{\bf Basis/Principles} & \textbf{Algorithms} \\
	\hline
	 &  \multicolumn{2}{|c|}{} & TAG~\cite{Madden:2002p2402}, DAG~\cite{Motegi:2005p1017}, \\
	 &  \multicolumn{2}{|c|}{\em Hierarchic} & I-LEAG~\cite{Birk:2006p3907}, \\
	 &  \multicolumn{2}{|c|}{} & Tributary-Delta~\cite{Manjhi:2005p7423}, \\
	&  \multicolumn{2}{|c|}{} & (Chitnis et al., 2008)~\cite{Chitnis:2008p6744}\\ 
	\cline{2-4}
	 & \multicolumn{2}{|c|}{ } & Push-Sum Protocol~\cite{Kempe:2003p700},\\
	\multicolumn{1}{|c|}{\bf Decomposable} & \multicolumn{2}{|c|}{ } &  Push-Pull Gossiping~\cite{Jelasity:2005p664},\\
	\multicolumn{1}{|c|}{\bf Functions} & \multicolumn{2}{|c|}{\em Averaging} & DRG~\cite{JenYeuChen:2006p3916},\\
	\multicolumn{1}{|c|}{\bf Functions} & \multicolumn{2}{|c|}{ } & Flow Updating~\cite{Jesus:2009p8366,Jesus:2010p9117},\\
	& \multicolumn{2}{|c|}{ } & (Chitnis et al., 2008)~\cite{Chitnis:2008p6744}\\
	\cline{2-4}
	 & \multicolumn{2}{|c|}{} & Sketches~\cite{Considine:2004p676},\\
	\multicolumn{1}{|c|}{} & \multicolumn{2}{|c|}{\em Sketches} & RIA-LC/DC~\cite{YaoChungFan:2008p5364,Fan:2010p9070},\\
	\multicolumn{1}{|c|}{} & \multicolumn{2}{|c|}{} & Extrema Propagation~\cite{Baquero:2009p7919},\\
	\multicolumn{1}{|c|}{} & \multicolumn{2}{|c|}{ } & Tributary-Delta~\cite{Manjhi:2005p7423}\\
    \hline
	\hline
	 & \multicolumn{2}{|c|}{} & Q-Digest~\cite{Shrivastava:2004p7854},\\
	\multicolumn{1}{|c|}{\bf Complex} & \multicolumn{2}{|c|}{\em Digests} & Equi-Depth\cite{Haridasan:2008p8141},\\
	\multicolumn{1}{|c|}{\bf Functions} & \multicolumn{2}{|c|}{ } & Adam2~\cite{Sacha:2010p9280}\\
    \hline
	\hline
	\cline{2-4}
	 &  \multicolumn{2}{|c|}{ } & Random Tour~\cite{Massoulie:2006p4521}, \\
	 &  \multicolumn{2}{|c|}{ } & Randomized Reports~\cite{Bawa:2003p673}, \\
	 &  \multicolumn{2}{|c|}{ }  & Sample \& Collide~\cite{Ganesh:2007p745,Massoulie:2006p4521},\\
	\multicolumn{1}{|c|}{\bf Counting} & \multicolumn{2}{|c|}{\em Sampling} & Capture-Recapture~\cite{Mane:2005p659},\\
	 & \multicolumn{2}{|c|}{ }  & Hop-Sampling~\cite{Kostoulas:2005p682,Kostoulas:2007p9773}, \\
	&  \multicolumn{2}{|c|}{ }  & Interval Density~\cite{Kostoulas:2005p682,Kostoulas:2007p9773},\\
	 &  \multicolumn{2}{|c|}{ }  & (Kutylowski et al., 2002)~\cite{Jurdzinski:2002p668}\\
	 &  \multicolumn{2}{|c|}{ } & (Horowitz and Malkhi, 2003)~\cite{Horowitz:2003p672}\\
	\hline
\end{tabular}
\caption{Taxonomy from the computation perspective.}
\label{tab:tax_computation}
\end{table}

Besides determining the supported aggregation function, the computational technique on which an aggregation algorithm is based constitutes a key element to define its behavior and performance, especially in terms of accuracy and reliability. \emph{Hierarchical} approaches are accurate and efficient (in terms of message and computational complexity), but not fault tolerant. \emph{Averaging} schemes are more reliable and also relatively accurate (converge along time), although less efficient (require more message exchanges). Approaches based on the use of \emph{sketches} are more reliable than hierarchical schemes, adding some redundancy and providing fast multi-path data propagation, however they introduce an approximation error (depending on the number of inputs and size of the used sketch). \emph{Digests} essentially consists on the reduction (compression) of all inputs into a fixed size data structure, using probabilistic methods and losing some information. Consequently, digests provide an approximation of the computed aggregation function, not the exact result. \emph{Sampling} schemes are also based on probabilistic methods to compute the \textsc{count}, 
being inaccurate and lightweight in terms of message complexity, as only a portion of the network is asked to participate. 

In the following sections, the main principles and characteristics of these distinct classes are explained in a comprehensive way, and some important examples are described. A taxonomy of the identified computational principles is displayed in Table \ref{tab:tax_computation}, associating them to the most relevant distributed aggregation algorithms.

\subsubsection{Hierarchical} 
\label{ssub:hierarchical}
\emph{Hierarchical} approaches take direct advantage of the decomposable property of some aggregation functions. Inputs are divided into separated groups and the computation is performed distributively in a hierarchical way. 
Algorithms from this class depend on the previous creation of a hierarchic communication structure (e.g. tree, clusters hierarchy), 
where nodes can act as \emph{forwarders} or \emph{aggregators}. Forwarders simply transmit the received inputs to an upper level node. Aggregators apply the target aggregation function directly to all received input (and its own), and forward the result to an upper level node. The correct result is yield at the top of the hierarchy, being the aggregation process carried out from the bottom to the top.

Algorithms from this class allow the computation of any decomposable function, providing the exact result (at a single node) if no faults occur. The global processing and memory resources required are equivalent to the ones used in a direct and centralized application of the aggregation function, but distributed across the network. However, these algorithms are not fault tolerant, e.g. a single point of failure may lead to the lost of all data beneath it.

Most of the algorithms from this category correspond to the ones belonging to the hierarchic communication class, like TAG~\cite{Madden:2002p2402}, DAG~\cite{Motegi:2005p1017}, and I-LEAG~\cite{Birk:2006p3907}. Other algorithms can be found combining a hierarchical computation with another computation principle, namely: Tributary-Delta~\cite{Manjhi:2005p7423} mix a common hierarchical computation, with the use sketches in regions close to the \emph{sink}; (Chitnis et al., 2008)~\cite{Chitnis:2008p6744} performs hierarchic aggregation on the top of groups, and averaging is applied inside each one. See sections \ref{sec:tree_based} and \ref{sec:hybrid} for more details about the aforementioned algorithms.

\subsubsection{Averaging}
\label{sec:averaging}

The \emph{Averaging} class essentially consists on the iterative computation of partial aggregates (averages), continuously averaging and exchanging data among all active nodes that will contribute  to the obtention of the final result. This kind of approach tends to be able to reach a high accuracy, with all nodes converging to the correct result along the execution of the algorithm. A typical application of this method can be found in most gossip-based approaches (section \ref{sec:gossip_based}), where all nodes continuously distribute a share of their value (averaged from received values) with some random neighbor, converging along time to the global network average (correct aggregation result). Algorithms from this category are more reliable than hierarchic approaches, working independently from the supporting network topology and producing the result at all nodes. However, they must respect an important principle, commonly designated as ``mass conservation'' in order to converge to the correct result. This invariant states that the sum of the aggregated values of all network nodes must remain constant along time~\cite{Kempe:2003p700}.

Algorithms based on this technique are able to compute decomposable and duplicate-sensitive functions, which can be derived from the average operation; using different inputs initializations (e.g. \textsc{count}), or combining functions executed concurrently (e.g. \textsc{sum}, obtained by multiplying the results from an average and a count). In terms of computational complexity, this method usually involves the computation of simple arithmetic operations (i.e. addition and division), using few computational resources (processor and memory) and being fast to execute (at each node). This kind of algorithms are able to produce (almost) exact results, depending on their execution time (without failures). The minimum execution time required by these algorithms (number of iterations), to achieve a high accuracy, is influenced by the network characteristics (i.e. size, connection degree, and topology) and the communication pattern used to spread the partial averages. The robustness of this type of aggregation algorithm is strongly related with their ability to conserve the global ``mass'' of the system. The loss of a partial aggregate (``mass'') due to a node failure or a message loss introduces an error, resulting in the subtraction of the lost value from the initial global ``mass'' (leading to the non-contribution of the lost amount to the calculation of the final result, and therefore to the convergence to an incorrect value). In this kind of methodology, it is important to enforce the ``mass'' conservation principle, assuming itself as a main invariant to ensure the algorithms correctness.

\paragraph{Push-Pull Gossiping}
The push-pull gossiping~\cite{Jelasity:2004p662} algorithm performs an averaging process, and it is gossip-based like the \emph{push-sum protocol~\cite{Kempe:2003p700}} (previously described in section \ref{sec:gossip_based}). The main difference of this scheme relies on the execution of an anti-entropy aggregation process. The concept of anti-entropy in epidemic algorithms consists in the regular random selection of another site, to resolve all the differences between the two, exchanging complete databases~\cite{Demers:1987p1534}. In particular, this algorithm executes an epidemic protocol to perform a pairwise exchange of aggregated values between neighbor nodes. Periodically, each node randomly chooses a neighbor to send its current value, and waits for the response with the value of the neighbor. Then, it averages the sent and received value, and calculates the new estimated value. Each time a node receives a value from a neighbor, it sends back its current one and computes the new estimate (average), using the received and sent values as parameters. In order to be adaptive and handle network changes (nodes joining/leaving), the authors consider the extension of the algorithm with a restarting mechanism (executing the protocol during a predefined number of cycles, depending on the desired accuracy). However, they do not address the ``mass'' conservation problem -- impact of message losses or node failures.

A further study of this aggregation algorithm is discussed in~\cite{Jelasity:2005p664}, proposing a more mature solution that covers some practical issues: split the algorithm execution in two distinct threads; use of timeouts to detect possible faults, ignoring data exchanges in those situations; suggest different versions of the algorithm according to the aggregation function to compute; suggest the execution of several instances of the algorithm in parallel to increase its robustness. 

\paragraph{DRG (Distributed Random Grouping)}
This approach~\cite{JenYeuChen:2006p3916} essentially consists on the continuous random creation of groups across the network, in which aggregates are successively computed (averaged). DRG was designed to take advantage of the broadcast nature of wireless transmission, where all nodes within radio range will be prone to ear a transmission, directing its application to WSN. The algorithm defines three different working modes for each node: \emph{leader}, \emph{member}, and \emph{idle} mode. According to the defined modes and the performed state transitions, the execution of the algorithm can be separated in three main steps. First, each node in idle mode independently decides to become a group leader (according to a predefined probability), and consequently broadcast a Group Call Message (GCM) to all its neighbors, subsequently waiting for members. Second, all nodes in idle mode which received a GCM from a leader respond to the first one with a Joining Acknowledgment (JACK) tagged with their aggregated value, becoming members of that group (updating their state mode accordingly). In the third step, after gathering the group members values from received JACKs, the leader computes (averages) the group aggregate and broadcast a Group Assignment Message (GAM) with the result, returning to idle mode. Each group member waits until it receives the resulting group aggregate from the leader to update its local value (with the one assigned in the GAM) and returns to idle mode, not responding to any other request until then. 

The repeated execution of this scheme -- creation of distributed random groups to perform in-group aggregation -- allows the eventual convergence of the estimate produced at all nodes to the correct aggregation result, as long as the groups overlap along time. The performance of DRG is influenced by the predefined probability of a node becoming leader, which determines its capacity to create groups (quantity and size of groups). Note that, in order to account for the occurrence of faults and avoid consequent deadlock situations that could arise in this algorithm, it is necessary to consider the definition of some timeouts (for the leaders to wait for JACKs, and the members to wait for a GAM). Intuitively, one will notice that the values set for these timeouts will highly influence the performance of the algorithm, although this detail is not addressed by the authors. An analysis of DRG on WSN with randomly changing graphs (modeling network dynamism) is provided in~\cite{JenYeuChen:2008p9361}, assuming that the graph only changes at the beginning of each iteration of the algorithm (unrealistic assumption in practice, otherwise this leads to mass loss).

\paragraph{Flow Updating} 
Flow Updating~\cite{Jesus:2009p8366} is a recent aggregation technique which is inspired from the concept of network flows (from graph theory). Unlike common averaging approaches, that start with the initial input value and iteratively change it by exchanging ``mass'' along the execution of the algorithm, this approach keeps the initial input unchanged, exchanging and updating flows associated to neighbors. The key idea is to explore the concept of flow, and instead of storing the current average at each node in a variable compute it from the input value and the contribution of the flows along the edges to the neighbors. In a sense, flows represent the value that must be transferred between two adjacent nodes for them to produce the same estimate, and are skew symmetric (i.e. the flow value from $i$ to $j$ is the opposite from $j$ to $i$: $f_{ij} = -f_{ji}$). For example: considering two directly connected nodes $i$ and $j$ with initial input values $v_i=1$ and $v_j=3$, for them to produce the same average $\frac{1+3}{2}=2$, the flows at node $i$ and $j$ must be respectively set to $f_{ij}=-1$ and $f_{ji}=1$.

In more detail, each node $i$ stores a flow value $f_{ij}$ to each neighbor $j$, besides its input value $v_i$ which is unchanged by the algorithm. Periodically, each node computes a new estimate $e_i'$ by averaging the ones received from neighbors $e_j$ with its own $e_i$ (obtained from subtracting all flows to its input value). The flows $f_{ij}$ to each neighbor $j$ are then locally updated in order to produce the new estimation result, adding the difference between the new estimate and the one previously received to the respective flow value. Afterwards, the node sends in a message the flows $f_{ij}$ to each neighbor $j$, as well as the new estimate. Upon reception of a message, node $j$ updates its flow $f_{ji}$ with the symmetric value addressed to him $-f_{ij}$, and keeps the received estimate to further compute the next average. The iterative execution of this process across the whole network allows the estimate of all nodes to converge to the global average of the input values.

This approach distinguishes itself from the existing averaging algorithms by its fault-tolerant capabilities. 
It solves the mass conservation problem observed on other averaging approaches when subject to message loss, that affect their correctness leading them to converge to a wrong value. Other approaches require additional mechanism to detect and restore the lost mass which is not feasible in practice. In contrast, Flow Updating is by design able to support message loss, only delaying convergence without affecting the convergence to the correct value, without requiring any additional mechanism. This is achieved by keeping the input values unchanged and performing idempotent flow updates which guarantee their skew symmetric property. Moreover, it has been recently shown that this approach is resilient to node crash and able to support churn (with requiring protocol restarts), self-adapting to network changes~\cite{Jesus:2010p9117}.

\paragraph{Other Approaches}

A well-known averaging approach, the Push-Sum (push-synopses) Protocol~\cite{Kempe:2003p700} has already been described in section \ref{sec:gossip_based}. In the last years, other approaches inspired by the Push-Sum Protocol have been proposed, intending to be more efficient in term of performance and robustness. Kashyap et al.~\cite{Kashyap:2006p775} reduces the number of messages needed (communication overhead) to compute an aggregation function at the cost of an increase in the number of rounds. G-GAP (Gossip-based Generic Aggregation Protocol)~\cite{Wuhib:2007p832} extends the \emph{push-synopses protocol}~\cite{Kempe:2003p700} to support discontinuous failures (no adjacent node can fail within a period of 2 rounds) by restoring the mass loss resulting from failures (temporarily storing at each node previous data contributions). 

Dimakis et al.~\cite{Dimakis:2008p1008,Dimakis:2006p1013} propose an algorithm to improve the convergence time in random geometric networks. This scheme is similar to push-pull gossiping~\cite{Jelasity:2004p662}, differing on the peer selection methods. Instead of selecting a one-hop node as target of the averaging step, peers are selected according to their geographical location. In particular, a location is randomly chosen and the node closer to that local is selected. A greedy geographic routing process is used to reach the node at the target location, assuming that nodes known their own geographic location. 

Two averaging algorithms for asynchronous and dynamic networks are proposed in \cite{Mehyar:2007p2525}. The core of the proposed schemes is based on a pairwise update, similarly to the push-pull gossiping (although not referred to the authors), addressing practical concerns that arise in asynchronous settings. In the first proposed algorithm nodes implement a blocking scheme to avoid the interference of other nodes in the update step and guarantee mass conservation. Additionally, a deadlock avoidance mechanism is considered, by imposing a sender-receiver relation on each link based on node UIDs (Unique IDentifiers). An extension to the first algorithm is proposed to cope with churn. The blocking mechanism (maintaining the directed relationship between nodes) is removed, and an additional variable is used to account for changes of each neighbor. When a node leaves the network, all its neighbors subtract the value associated to it from their state.

\subsubsection{Sketches} 
\label{sec:sketches}

The main principle of this kind of aggregation algorithm is based on the use of an auxiliary data structure with a fixed size, holding a \emph{sketch} of all network values. The input values are used to create sketches that are aggregated across the network, using specific operations to update and merge them. Operations on sketches are order and duplicate insensitive, enabling them to be aggregated through multiple paths, being independent from the routing topology. This kind of technique is based on the application of a probabilistic method, generally allowing the estimation of the sum of the values held in the sketch.

Sketching techniques can be based on different methods, with different accuracy bounds and computational complexities. Algorithms from this class are mostly based on the application (with some improvements) of two main ideas: \emph{hash sketches}~\cite{Flajolet:1985p2833,Whang:1990p8589,Durand:2003p8548,Flajolet:2007p8568} and \emph{k-mins sketches}~\cite{Cohen:1997p8473}. 

Hash sketches allow the probabilistic counting of the number of distinct elements in a multiset (cardinality of the support set). This type of sketch essentially consists in a map of bits, initially set to zero, where each item is mapped into a position in the binary valued map (generally involving a uniform hashing function) setting that bit to one. The distinct count is estimated by checking the position of the most significant one bit (leftmost), or counting the number of zero bits in the sketch. The first hash sketching technique was proposed by Flajolet and Martin~\cite{Flajolet:1985p2833}, being commonly designated as FM sketches (uniformly hash items into an integer, and maps only the less significant one bit of its bitmap representation to the sketch). In this first study, the authors also proposed the PCSA (Probabilistic Counting with Stochastic Averaging) algorithm to reduce the variance of the produced estimate, using multiple sketches and averaging their estimate (distributing the hash of an element to only one of the sketches). Another approach, Linear Counting~\cite{Whang:1990p8589} uses a hash function to directly map each element into a position of the sketch (setting that bit to one), and use the count of the number of zeros to produce an estimate. A further improvement to PCSA, designated LogLog, was described in \cite{Durand:2003p8548}, reducing required memory resources (an optimized version super-LogLog is also proposed, improving accuracy and optimizing memory usage applying a truncation and restriction rule). HyperLogLog~\cite{Flajolet:2007p8568} recently improved LogLog, consuming less memory to achieve a matching accuracy. 

The k-mins sketches method was first introduced to determine the size of the transitive closure in directed graphs~\cite{Cohen:1997p8473}. It consists on assigning $k$ independent random ranks to each item according to a distribution that depends on its weight, and keeping in a vector of the minimum ranks in the set. The obtained $k$-vector of the minimum ranks is used by an estimator to produce an approximated result. In other words, it can be said that k-mins sketches reduces the estimation of the sum to the determination of minimums of a collection of random numbers (generated using the sum operands as input parameters of the random distribution from which they are drawn). An improved alternative to k-mins sketches, designated bottom-k sketches, was recently proposed in \cite{Cohen:2007p8379}.

The computational cost of sketching is dependent on the complexity of the operations involved in the creation and update of the sketches (e.g. hashing functions, random number generation, minimum/maximum determination), and the resources used by the estimator to produce a result. Algorithms based on sketches are not accurate, being based on probabilistic methods and introducing an error factor in the computed aggregation function. There is a trade-off between the accuracy and the size of the sketches. The greater the sketch size the tighter are the accuracy bounds of the produced estimate, although requiring additional memory resources and a larger processing time. This kind of aggregation algorithm tends to be fast, although conditioned by the dissemination protocol used to propagate the sketches, being able to produce an approximate result after a number of iterations close to the minimum theoretical bound (the network diameter).

\paragraph{RIA-LC/DC}
Fan and Chen~\cite{YaoChungFan:2008p5364} proposed a multi-path routing aggregation approach for WSN based on the use of Linear Counting (LC) sketches~\cite{Whang:1990p8589}, which they later named RIA-LC (Robust In-network Aggregation using LC-sketches)~\cite{Fan:2010p9070}. The algorithm proceeds in two phases, like common multipath hierarchy-based approaches (see Section \ref{sec:tree_based}). In the first phase, the aggregation request (query) is spread from the sink throughout the whole network, creating a multipath routing hierarchy. In the second phase, starting at the lower level of the hierarchy, nodes respond to the aggregation request by creating a LC-sketch correspondent to its current local readings and sending it to the nodes at the upper level. All received sketches are combined with the local one (using the OR operation), and the result is sent to the next level until the top of the hierarchy is reached where the sink computes the aggregation estimate from the resulting LC-sketch.

Equation \ref{eq:lc-sketch} is used to estimate the number of distinct items represented in a LC-sketch, where $m$ is the size of the allocated bit vector, and $z$ is the count of the number of bits with value equal to zero. In order to allow the computation of the \textsc{sum}, each node creates a sketch by mapping a number of distinct items corresponding to its input value. For example, assuming that each node has a unique ID, if the node $i$ has an input equal to 3, it maps the items $(ID_i,1)$, $(ID_i,2)$, and $(ID_i,3)$ into the LC-sketch. In more detail, in this case the use of an hash function from the original LC-sketch design (to map duplicated items to the same bit) is replaced by a uniform random generator (since there are no duplicate items), randomly setting to 1 a number of bits equal to the input value.

\begin{equation}\label{eq:lc-sketch}
\hat{n} = -m\ln{(z/m)}
\end{equation}

The authors show by theoretical comparison and experimental evaluation that their approach outperforms in terms of space and time requirements the ones based on FM sketches~\cite{Flajolet:1985p2833}, namely Sketches~\cite{Considine:2004p676} (see \ref{sec:tree_based}). They also claim a higher accuracy and lower variance when compared with existing sketch schemes. Moreover, they tackle some practical issues, like message size constraints, avoid the use of hash functions, and enable the specification of an approximation error.

Recently, the authors improved RIA-LC by considering the use of sketches with variable sizes instead of fixed size sketches, referring to the new technique as RIA-DC (Robust In-network Aggregation using Dynamic Counting sketches)~\cite{Fan:2010p9070}. The authors observed that the large preallocated sketches used in RIA-LC were wasting space, since at the beginning of the computation must of the bits are set to zero. In RIA-DC the initial size of sketches is variable and depends on the local sensor reading. Along the aggregation process the size of the sketches is adjusted (gradually increasing toward the sink), in order to satisfy a given accuracy constrain. RIA-DC decreases message overhead and energy consumption compared to RIA-LC, keeping similar accuracy properties.

\paragraph{Extrema Propagation}
\label{par:extrema}
This approach reduces the computation of an aggregation function, more precisely the sum of positive real numbers, to the determination of the minimum (or maximum) of a collection of random numbers~\cite{Baquero:2009p7919,Baquero:vu}. Initially, a vector $x_i$ of $k$ random number is created at each network node $i$. Random numbers are generated according to a known random distribution (e.g., exponential or gaussian), using the node initial value $v_i$ as the input parameter for the random generation function (e.g., as the rate of an exponential distribution). Then, the execution of the aggregation algorithm simply consists of the computation of the pointwise minimum (or alternatively maximum) between all exchanged vectors. This technique supports the use of any information spreading algorithm as a subroutine to propagate the vectors, since the calculation of minimums is order and duplicate insensitive. In particular, the authors consider that at each round all nodes send their resulting vector to all their neighbors.

At each node, the obtained vector is used as a sample to produce an approximation of the aggregation result, applying a maximum likelihood estimator derived from extreme value theory (branch of statistics dealing with the extreme deviation from the median of a probabilistic distribution). For example, considering the generation at each node of $k$ random number with an exponential distribution of rate $v_i$ and the use of the minimum function to aggregate the vectors. Equation~\ref{eq:extrema} gives the estimator for the \textsc{sum} of all $v_i$ from the sample of minimums $x_i[1],...x_i[k]$ in the vector $x_i$, with variance $\textsc{sum}^2/(k-2)$:

\begin{equation}\label{eq:extrema}
\widehat{\textsc{sum}} = \frac{k-1}{\sum_{j=1}^{k}x_i[j]}
\end{equation}

This algorithm is focused on obtaining a fast estimate, rater than an accurate one. Although, the accuracy of this aggregation algorithm can be improved by using vectors of larger size, adjusting $k$ to the desired relative accuracy (e.g. $k=387$ for a maximum relative error of $10\%$, with a confidence of $95\%$). A further extension to the protocol to allow the determination of the network diameter has been proposed in ~\cite{Cardoso:2009p7920}.

\paragraph{Other Approaches}

A representative approach based on FM sketches has already been described in section \ref{sec:tree_based} -- Sketches~\cite{Considine:2004p676}. In this multi-path approach, a generalization of PCSA is used to distinguish the same aggregates received from multiple paths, and subsequently manage to compute duplicate-sensitive aggregation functions. Other similar approaches can be found in the literature based on hash sketches, like \emph{Synopsis Diffusion}~\cite{Nath:2004p1114} and Wildfire~\cite{Bawa:2004p7601}. These approaches apply essentially the same aggregation process, operating in two phases (request/response) and only differing on small aspects.

\emph{Synopsis Diffusion}~\cite{Nath:2004p1114} is an aggregation approach for WSN close to the one proposed by Sketches~\cite{Considine:2004p676}. In a sense, this work presents a more generic framework relying on the use of duplicate insensitive summaries (i.e. hash sketches), which they call ODI (Order- and Duplicate-Insensitive) synopses. Namely, they generically define the synopses functions (i.e., generation, fusion and evaluation) required to compute aggregation functions, and provide examples of ODI synopses to compute more ``complex'' aggregates (i.e., not decomposable aggregation functions). For instance, besides the scheme based on FM sketches, they propose other data structures (and respective functions) to uniformly sample sensor readings and compute other sampling based aggregation functions. The authors also tackled additional practical concerns. Namely, they explored the possibility of implicitly acknowledging ODI synopses to infer messages losses, and suggested simple heuristics to modify the established routing topology (assigning nodes to another hierarchic level), in order to reduce loss rate. 

Wildfire~\cite{Bawa:2004p7601} is based on the use of FM sketches to estimate \textsc{sum}, but it is targeted for dynamic networks. Despite the fact of operating in two phases like previous hash sketch approaches, unlike them it does not establish any specific routing structure (i.e. multipath hierarchy) to aggregate sketches. After receiving the query, nodes start combining the received sketches with their current one, and then send the result if it differs from the previous one.

A distributed implementation of some basic hash sketches schemes has been proposed in~\cite{Ntarmos:2006p7363,Ntarmos:2009p7779}. DHS (Distributed Hash Sketches) is supported by a DHT, taking advantage of the load balancing properties and scalability of such structure. 
More specifically, the authors describe how to build DHS based on PCSA~\cite{Flajolet:1985p2833} and supper-LogLog~\cite{Durand:2003p8548}. 

Mosk-Aoyama and Shah~\cite{MoskAoyama:2008p5417,MoskAoyama:2006p686} proposed an algorithm, called COMP, to compute the sum of values from individual functions (referred to as separable functions). This algorithm is very similar to \emph{Extrema Propagation} but less generic, as it is restricted to the properties of exponential random variables distribution. Furthermore, COMP uses a biased estimator, being less accurate than Extrema Propagation which uses unbiased ones. 


\subsubsection{Digests} 
\label{ssub:digests}

This category includes algorithms that allow the computation of more complex aggregation functions, like quantiles (e.g., median) and frequency distributions (e.g., mode), besides common aggregation functions (e.g., count, average and sum). Basically, algorithms from this class produce a \emph{digest} that summarizes the system data distribution (e.g., histogram). The resulting \emph{digest} is then used to approximate the desired aggregation functions. We refer to a \emph{digest} as a data structure with a bounded size, that holds an approximation of the statistical distribution of input values in the whole network. 
This data structure commonly corresponds to a set of values or ranges with an associated counter. 

Digests provide a fair approximation of the data distribution, not holding an exact representation of all the system values due to efficiency and scalability reasons. The accuracy of the result obtained from a digest depends on its quality (i.e., used data representation) and size. Digest allow the computation of a wider range of aggregation functions, but usually require more resources and are less accurate than the other more specialized approaches.

\paragraph{Q-Digest} 
\label{par:q_digest}

An aggregation scheme that allow the approximation of complex aggregation functions in WSN is proposed in \cite{Shrivastava:2004p7854}. This approach is based on the construction and dissemination of q-digests (quantile digests) along a hierarchical routing topology (without routing loops and duplicated messages). A q-digest consists of a set of buckets, hierarchically organized, and their corresponding count (frequency of the values contained by the bucket). Buckets are defined by a range of values $[a, b]$ and can have different sizes, depending on the distribution of values they represent. Each node maintains a q-digest of the data available to it (from its children). Q-digests are built in a bottom-up fashion, by merging received digests from child nodes, and further compressing the resulting q-digest according to a specific compression factor (less frequent values are grouped in large buckets). Aggregation functions are computed by manipulating (e.g., sort q-digest nodes) and traversing the q-digest structure according to a specific criteria (depending on the function to be computed). 

The authors provide an experimental evaluation, where they show that q-digests allow the approximation of quantile queries using fixed message sizes, 
saving bandwidth and power when compared to a naive scheme that collects all the data. The naive scheme obtains an exact result, but with increasing message size along the routing hierarchy. Obviously, there is a trade-off between the obtained accuracy and the message size used. The authors suggest a way to compute the confidence factor associated to a q-digest (i.e., error associated to a query), but the effect of faults is not considered in their study.

\paragraph{Equi-Depth} 
\label{par:equi_depth}

A gossip-based approach to estimate the network distribution of values is described in \cite{Haridasan:2008p8141}. This scheme is based on the execution of a gossip protocol and the application of specific merge functions to the exchanged data, to restrict storage and communication costs. In more detail, each node keeps a list of $k$ values (digest), initially set to its input value. At each round, nodes get the list of values from a randomly chosen neighbor and merge it with their own, applying a specific procedure. The result from the execution of several rounds produces an approximation of the network distribution of values (i.e., histogram). Four merging techniques were considered and analyzed by the authors: \emph{swap}, \emph{concise counting}, \emph{equi-width histograms}, and \emph{equi-depth histograms}.

Swap simply consists in randomly picking $k$ values from the two lists (half from each of them) and discarding the rest. Although simpler, by discarding half of the available data in each merge, important information is likely to be lost.

Concise counting associates a tuple, value and count, to each list entry. The merge process consists in sorting the tuples (by value), and individually merging the tuples with the closest values, in order to keep a fixed list size. Tuples are merged by randomly choosing one of the values and adding their count.

The equi-width technique breaks the range of possible values into bins of equal size, associating a counter to each one. Initially, nodes consider the range from 0 to the current input value, as the extremes are not known. Bins are dynamically resized when new extremes are found: all bins are mapped into larger ones, based on their middle value and the range of the new bin, adding their counter to the new mapped bin. This technique requires only the storage of the extreme values and counts, since all bins have an equal width, reducing the volume of data that needs to be stored and exchanged when compared to other techniques (e.g., concise counting). However, equi-width can provide very inaccurate results for severely skewed distributions.

In equi-depth, bins are divided not to be of the same width but to contain approximately the same count. Initially, fixed size bins are set, each represented by a pair $<$value, counter$>$, dividing the range from 0 to the input value. Whenever data is exchanged, all pairs (received and owned) are ordered, and consecutive bins that yield the smallest combined bins (in terms of count) are merged, repeating the process until the desired number of bins is obtained. Bin merge consists in adding the counters and using the arithmetic weighted mean as value. This method intends to minimize the counting disparity across bins.

In order to deal with changes in input values along time, the authors consider the execution of the protocol in phases, restarting it. The authors experimentally evaluated their protocol comparing the previous merging techniques. The results obtained show that equi-depth outperformed the other approaches, providing a consistent trade-off between accuracy and storage requirements for all tested distributions. The author also evaluated the effect of duplicates, from the execution of the gossip protocol. They argue from the results obtained that although duplicates bias the estimated result, it is more advantageous (simpler and efficient) to assume their presence than to try to remove them. The occurrence of faults and change in the input values were not evaluated.

\paragraph{Adam2} 
\label{par:adam2}
Adam2 is a gossip based algorithm to estimate the statistical distribution of values across a decentralized system~\cite{Sacha:2010p9280}. More precisely, this scheme approximates the CDF (Cumulative Distribution Functions) of an attribute, which can then be used to derive other aggregates. In this case, a ``digest'' is composed by a set $H_i$ of $k$ pairs of values $(x_k,f_k)$, where $x_k$ represents an interpolation point and $f_k$ is the fraction of nodes with value less or equal than $x_k$. At a high abstraction level, it can be said that the algorithm simply executes several instances of an averaging protocol (i.e., Push-Pull Gossiping~\cite{Jelasity:2005p664}) to estimate the fraction of nodes in each pair of the CDF. 

In more detail, each node can decide to start an instance of Adam2 according to a predefined probability $\frac{1}{\hat{n}_iR}$, where $\hat{n}_i$ is the current network size estimate at node $i$ and $R$ is an input parameter that regulates the aggregation instances frequency (i.e. on average one every $R$ rounds). Each instance is uniquely identified by its starting node. Initially, the starting node $i$ initializes the interpolation set $H_i$ in the following way: fractions $f_k$ are set to $1$ if the node attribute reading $v_i$ is less or equal than the corresponding interpolation value $x_k$, and set to $0$ otherwise. Nodes store a set of interpolation points $H_i$ for each running algorithm instance (initiated by a node $i$). Upon learning about a new instance, a node $j$ initializes its $H_i$ setting $f_k = 1$ if $a_j \leq x_k$ and $f_k = 0$ otherwise, and starts participating in the protocol. A push-pull like aggregation is then performed, where nodes randomly choose a neighbor to exchange their set $H_i$, which are subsequently merged by averaging the fractions at each interpolation point. Along time, the fractions will eventually converge at each node to the correct result associated to each pair. After a predefined number of rounds (\emph{time-to-live}) the CDF is approximated by interpolating the points of the resulting set $H_i$. Note that, Adam2 concurrently estimates (by averaging) other aggregation functions besides \textsc{CDF}, namely \textsc{count} to determine the network size, and \textsc{min}/\textsc{max} to find the extreme attribute values. The result from these aggregation functions are later used as input values to the next instances of the algorithm to tune and optimize its execution (i.e., calculate the instance starting probability, and set new interpolation points).

Like in Push-Pull Gossiping~\cite{Jelasity:2004p662,Jelasity:2005p664}, Adam2 handles dynamism (i.e., attribute changes and churn) by continuously starting new instances of the algorithm -- restart mechanism. The authors evaluated the algorithm by simulation, comparing it with previous techniques to compute complex aggregates (e.g., \emph{Equi-Depth}). The results obtained show that Adam2 outperforms the compared approaches, exhibiting better accuracy.

\paragraph{Other Approaches} 
\label{par:other_approaches}

One of the first algorithms to compute complex aggregation functions in WSN was introduced by Greenwald and Khanna~\cite{Greenwald:2004p7935}. Their approach is similar to the one previously described for q-digest (\ref{par:q_digest}): nodes compute quantile summaries (digest) that are merged in a bottom-up fashion along a tree topology, until the root is reached.

Another gossip based scheme to estimate the distribution of input readings, able to detect outliers, was introduced in \cite{Eyal:2009p8140,Eyal:2010p9926}. In a nutshell, this approach operates like the push-sum protocol~\cite{Kempe:2003p700} (described in Section \ref{sec:gossip_based}), but manipulates a set of clusters (digests) instead of a single value, applying a specific clustering procedure.

In general, existing aggregation approaches can be extended to compute more complex aggregation functions, for instance combining them with an additional sampling technique. However, this additional functionality is not part of the essence of their core algorithm, bearing different characteristics (e.g. accuracy) and concerns. Some examples can be found in~\cite{Kempe:2003p700} where push-sum is extended with a push-random protocol to obtain random samples, and in \cite{Cohen:2004p1156} which introduces algorithms to estimate several spatially-decaying aggregation functions.


\subsubsection{Counting} 
\label{sec:counting}

This category refers to a restricted set of distributed algorithms, designed to compute a specific aggregation function: \textsc{count}. \textsc{count} allows the determination of important properties in the design of some distributed applications. For instance, in this context it finds a common practical application in the determination of the size of the system (or group), or to count the number of votes in an election process. The algorithms from this class rely on the use of some randomized process,  
most of them usually based on the execution of some \emph{sampling} technique to provide a probabilistic approximation of the size of the sample population. Nonetheless, a few algorithms are found that do not explicitly collect samples for size estimation, instead applying a probabilistic \emph{estimator} over some observed events. 


Algorithms based on sampling are strongly influenced by the probabilistic method used to obtain the result, inheriting its properties. For instance, the accuracy of the algorithm corresponds to the one provided by the probabilistic method used, being bounded by the error factor associated with it. Several probabilistic methods have been applied to samples to yield a counting estimation, namely: \emph{birthday problem}~\cite{DasGupta:2005p3722} -- concerns the probability of two elements sampled out of a population not being repeated, inspired from the probability of two people out of a group not having a matching birthday; 
\emph{capture-recapture}~\cite{Schwarz:1999p691} -- probabilistic method based on the repeated capture of samples from a closed population (population that maintains a fixed size during the sampling process), where the number of common elements between samples are accounted to provide an estimate of the population size; \emph{fundamental probabilistic methods} -- application of Bernoulli based sampling methods~\cite{Cheng:2010p8605}, and other basic probabilistic concepts on some sampled statistical information, like the distances between nodes (number of hops) or the number of messages successfully sent/received, in order to estimate de size of the network. In all cases, typically sampling is performed at a single node, and it can take several rounds to collect a single sample. Moreover, an estimation error is always present, even if no faults occur. For example, in \emph{Sample \& Collide}~\cite{Ganesh:2007p745,Massoulie:2006p4521} the estimation error can reach $20\%$, and a sampling step takes $\bar{d}T$ (where $\bar{d}$ is the average connection degree and $T$ is a predefined timer that must be sufficiently large to provide a good sample quality), needing to be repeated until $l$ sample collisions are observed.

As previously referred, in some cases a size estimation can be obtained by directly applying an estimator on some available system knowledge (observed events or other known properties), without any previous explicit sampling. Although, in general the estimator inputs result from other sampling sources. For instance, in the approach proposed by Horowitz and Malkhi~\cite{Horowitz:2003p672} (see section~\ref{sec:ring}) an estimator function is used at each node  to estimate the network size, based on the observation of two events (nodes joining or leaving the network), incrementing/decrementing the estimator. In this case the nodes joining/leaving at each node can be seen as the input sample used to provide the estimate. Other approaches, like the one proposed in \cite{Dolev:2003p7204,Dolev:2006p7210}, provide a size estimation based on knowledge of the routing structure, in this particular case counting the number of high degree nodes (which can be considered the input sample). This kind of techniques does not provide accurate results, in most cases yielding a rough approximation to the correct value.

%
%

\paragraph{Sample \& Collide}
This approach~\cite{Ganesh:2007p745,Massoulie:2006p4521} addresses the problem of counting the number of peers in a P2P overlay network, inspired by a birthday problem technique (first proposed by Bawa et al. on a technical report~\cite{Bawa:2003p673}).
The application of this probabilistic method requires the collection of uniform random samples. To this end, the authors proposed a peer sampling algorithm based on the execution of a continuous time random walk, in order to obtain unbiased samples (asymptotically uniform). The sampling routine proceeds in the following way: an initiator node $i$ sets a timer with a predefined value $T$, which is sent in a sampling message to a randomly selected neighbor; upon receiving a sampling message, the target node (or the initiator after setting the timer) picks a random number $U$ uniformly distributed within the interval $[0, 1]$, and decrements the timer by $\log{(1/U)}/d_i$ (i.e. $T \gets T - \log{(1/U)}/d_i$, where $d_i$ is the degree of node $i$); if the resulting value is less or equal than zero ($T \le 0$) then the node is sampled, its identification is returned to the initiator and the process stops; otherwise the sampling message is sent to one of its neighbors, chosen uniformly at random. The quality of the samples obtained (approximation to a uniform random sampling) depends on the value $T$ initially set to the timer. The described sampling step (to sample one peer) must be repeated until one of the nodes is repeatedly sampled a predefined number of times $l$ (i.e. $l$ sample collisions are observed). After concluding this sampling process, the network size $n$ is estimated using a Maximum Likelihood (ML) method. The ML estimate can be computed by solving Equation~\ref{eq:sample_collide1}, where $C_l$ corresponds to the total number of samples until one is repeated $l$ times, using a standard bisection search. Alternatively, the result can be approximated within $\sqrt{n}$ of the ML-estimator by Equation~\ref{eq:sample_collide2} (asymptotically unbiased estimator), which is computationally more efficient. 

\begin{equation}\label{eq:sample_collide1}
\sum_{i=0}^{ C_{l}-l-1} \frac{i}{n -1} - l = 0
\end{equation}
\begin{equation}\label{eq:sample_collide2}
\widehat{n} = C_{l}^{2}/2l
\end{equation}

The accuracy of the produced result is determined by the parameter $l$, and its fidelity depends on the capacity of the sampling method to provide uniformly distributed random samples ($T$ must be sufficiently large).

\paragraph{Capture-Recapture}
Mane et al.~\cite{Mane:2005p659} proposed an approach based on the capture-recapture statistical method to estimate the size of closed P2P networks (i.e. networks of fixed size, with no peers joining or leaving during the process). This method requires two or more independent random samples from the analyzed population, and further counting the number of individuals that appear repeated in each sample. The authors use random walks to obtain independent random samples. Considering a two-sample strategy, two random walks are performed from a source node, one in each sampling phase (capture and recapture). In more detail, each random walk proceeds in the following way: the source node sends a message to a randomly selected neighbor, which at its turn forwards the message to another randomly chosen neighbor; the process is repeated until a predefined maximum number of hops is reached (parameter: \emph{time-to-live}) or the message gets back to a node that has already participated in the current random walk. During this process, the information about the traversed path (i.e. the UIDs of all participating nodes) is kept in the forwarded message. When one of the random walk stopping criteria is met, the message is sent back to the source node with the list of the ``captured'' nodes, following the reverse traversed path (stored in the message). The information received at the source node from the sampling steps is used to compute the estimate $\widehat{n}$ of the network size, applying Equation \ref{eq:capture-recapture} (where $n_{1}$ is the number of nodes caught in the first sample, $n_{2}$ is the number of nodes caught in the second sample, and $n_{12}$ represent the number of recaptured nodes, i.e. caught in both samples).

\begin{equation}\label{eq:capture-recapture}
\widehat{n} = \frac{((n_{1}+1)\times(n_{2}+1))}{(n_{12}+1)}
\end{equation}

\paragraph{Hop-Sampling}
One of the approaches proposed by Kostoulas et al.~\cite{Kostoulas:2005p682,Kostoulas:2007p9773} to estimate the size of dynamic groups is based on sampling the receipt times (hop counts) of some nodes from an initiator. Receipt times are obtained across the group from a gossip propagation started by a single node, the initiator, that will further sample the resulting hop counts of some nodes to produce an estimate of the group size. 
In more detail, the protocol proceeds as following: the initiator starts the process by sending an \emph{initiating} message (to itself); upon receiving the initiating message nodes start participating in the protocol, forwarding it to a number (\emph{gossipTo}) of other targets, until a predefined number of rounds (\emph{gossipFor}) is exceeded, or a maximum quantity of messages (\emph{gossipUntil}) have been received; gossip targets are chosen uniformly at random from the available membership, excluding nodes in a locally maintained list (\emph{fromList}) from which a message has already been received; exchanged messages carry the distance to the initiator node, which is measure in number of hops; each node keeps the received minimum number of hops (\emph{MyHopCount}), and sends the current value incremented by one. 
After concluding the described gossip process, waiting for a predefined number of rounds (\emph{gossipResult}), the initiator samples the number of hops (\emph{MyHopCount}) from some nodes selected uniformly at random. The average of the sampled hop counts is then used to estimate the logarithm of the size of the group ($log{(n)}$).
Alternatively to the previous sampling process, where nodes wait for the initiator sample request, nodes can decide themselves to send their hop count value back to the initiator node, according to a predefined probability to allow only a reduced fraction of nodes to respond.

\paragraph{Interval Density}
A second approach to estimate the size of a dynamic group has been proposed in \cite{Kostoulas:2005p682,Kostoulas:2007p9773}. This algorithm measures the density of the process identifiers space, determining the number of unique identifiers within a subinterval of this space. The initiator node passively collects information about existing identifiers, snooping the information of complementary protocols running on the network. The node identifiers are mapped to a point in the real interval $[0, 1]$ by applying a hash function to each one. The initiator estimates the group size by determining the number of sampled identifiers $X$ lying in a subinterval $I$ of $[0, 1]$, returning $X/I$. Notice that this kind of approach assumes a uniformly random distribution of the identifiers, or uses strategies to reduce the existing correlation between them, in order to avoid biased estimations.

\paragraph{Other Approaches}
Some counting approaches based on a centralized probabilistic polling to collect samples were previously described in this work (in section~\ref{sec:flooding}). Namely, \emph{randomized reports} which illustrate the basic idea of probabilistic polling, and another approach \cite{Jurdzinski:2002p668} that samples the number of message successfully sent in a single-hop wireless network (further improved in \cite{Kabarowski:2006p694}).

Other probabilistic polling algorithms are also available in the specific context of multicast groups, to estimate their membership size. For example, in \cite{Friedman:1999p7356} some older mechanisms were analyzed and extended, and in \cite{Alouf:2002p7162} an algorithm using an estimator based on Kalman filter theory was proposed to estimate the size of dynamic multicast groups.


\section{Summary and Practical Guidelines} 
\label{sec:practical_guidelines}

Here, we summarize the properties of the main classes of algorithms, stating their advantages and disadvantages (see Table \ref{tab:main_agg_class_prop}), and give some guidelines about their use in specific settings.

Hierarchy-based approaches (see \ref{sec:tree_based} and \ref{ssub:hierarchical}) require a specific routing structure (e.g. spanning tree) to operate, and thus are limited by the ability of such structure to cope with churn and link failures. However, this kind of approach is very cheap in term of messages exchange, requiring only $2N-1$ messages\footnote{This is for scenarios where radio broadcast is used to transmit data, such as in WSN. Otherwise, approximately $\bar{d}(2N-1)$ messages are required, where $\bar{d}$ is the average degree.} to compute the correct average at the sink, i.e. two messages for each $N$ nodes (except the sink), one to broadcast the aggregation request to its child nodes and another to send the result to its parent. In terms of time, the aggregation process takes $2h$ rounds (at most $2D$, with $D$ representing the network diameter), where $h$ is the height of the routing hierarchy, i.e. $h$ rounds to spread the aggregation request and another $h$ rounds to aggregate the results from child nodes to parents. This kind of technique is commonly used in energy constrained environments (i.e. WSN), taking advantage of the reduced messages exchanges. Therefore, we would only recommend the application of such aggregation schemes to fault free scenarios, which is often not the case. This kind of approach can be significantly affected by the occurrence of a single failure, losing all the subtree data and greatly impacting the result produced at the sink. For this reason, in scenarios where faults might occur and without regarding energy efficiency, sketch techniques should be preferred, at least providing some path redundancy to reach the sink (at the cost of $k$ factor increase in terms of messages, with $k$ representing the number of alternative routing paths).

Sketches approaches (see \ref{sec:sketches}) can be applied independently from the underlying routing topology, being adequate to use in faulty scenarios where only a fair approximation of the aggregate is required. This kind of technique is fast, the closest to the theoretical minimum, requiring only $D$ rounds for all nodes to obtain the estimation result, achieving this at a total cost of $\bar{d}ND$ messages (i.e. each $N$ nodes send at most one message to each $d$ neighbors at each round)\footnote{In the case of WSN, with radio broadcast the total message cost is reduced to $ND$.}. This kind of approach is adequate for faulty scenarios, especially if one privileges obtaining a fast estimate rather than a precise one. In particular, we recommend the use of Extrema Propagation\cite{Baquero:2009p7919} (see \ref{par:extrema}) which is able to provide a better and unbiased estimation, when compared to other sketch algorithms. Nonetheless, if a precise estimation is required in a faulty environment, another type of aggregation approach should be chosen, namely an averaging technique. 

Averaging algorithms (see \ref{sec:gossip_based} and \ref{sec:averaging}) work independently from the routing topology, and have the particularity to converge along time to the correct result, being able to produce results at all nodes with high accuracies even in faulty environments. The execution time of this kind of algorithms depends on the target accuracy, converging exponentially with linear rounds 
(at an approximately constant convergence factor between each round), with all nodes sending from one to $d$ message at each round. This kind of approach is slower and consequently requires more messages (although smaller) than sketches, but can exhibit better properties in terms of fault-tolerance, especially to cope with churn. In particular, Flow Updating~\cite{Jesus:2010p9117} is the only known fault-tolerant aggregation approach which is able to continuously adapt to network changes, without resorting on any kind of restart mechanism like others (i.e. that periodically reset and start a fresh execution of the algorithm). Moreover, apart from Flow Updating, it has been shown that the other averaging approaches  exhibit some dependability issues, not converging to the correct value. For this reason, from the existing averaging approaches we will only recommend the use of Flow Updating, which is adequate to use in dynamic and faulty scenarios, where accurate estimates are required and there is no strict constraint on the quantity of exchanged messages.

Sampling aggregation techniques (see \ref{sec:random_walk} and \ref{sec:counting}) does not seem to bring any advantages when compared to the other kind of approaches. This kind of approach provides an irregular approximation at a single node, not being accurate and usually restricted to the computation of a single aggregation function: \textsc{count}. Furthermore, the random walk based approaches are usually slow, taking several rounds to obtain a sample, and unreliable as the random walk might be lost in a faulty environment. 

\begin{table}[t]
\centering
\begin{tabular}{c|l|l|l|}
\cline{2-4}
  & \multicolumn{1}{|c|}{\bf Advantage} & \multicolumn{1}{|c|}{\bf Disadvantage} & \multicolumn{1}{|c|}{\bf Requirements}\\
\hline
\multicolumn{1}{|c|}{} & - accurate & - result at a single  & - specific routing \\
\multicolumn{1}{|c|}{\bf Hierarchical} & \hspace{1ex} (without faults); &  \hspace{1ex} node;  & \hspace{1ex} structure (e.g. \\
\multicolumn{1}{|c|}{} & - very efficient &  - not fault-tolerant;  & \hspace{1ex} spanning tree);\\
\multicolumn{1}{|c|}{} & \hspace{1ex} (messages); &  & \\
\hline
\multicolumn{1}{|c|}{} &  &  &  - local knowledge\\
\multicolumn{1}{|c|}{} & - very fast; &  & \hspace{1ex} of neighbor IDs,\\
\multicolumn{1}{|c|}{\bf Sketches} & - result at all nodes; & - less accurate; & \hspace{1ex} or global UIDs;\\
\multicolumn{1}{|c|}{} & - fault-tolerant; &  & - source of\\
\multicolumn{1}{|c|}{} &  &  & \hspace{1ex} randomness;\\
\hline
\multicolumn{1}{|c|}{ } & - accurate; &  &  \\
\multicolumn{1}{|c|}{\bf Averaging} & - result at all nodes; & - less efficient & - local knowledge \\
\multicolumn{1}{|c|}{ } & - fault-tolerant; & \hspace{1ex} (messages); & \hspace{1ex} of neighbor IDs;\\
\multicolumn{1}{|c|}{ } & - churn support; &  & \\
\hline
\multicolumn{1}{|c|}{} &  & - not accurate & \\
\multicolumn{1}{|c|}{\bf Sampling} & - efficient & - result at a single & - global UIDs;\\
\multicolumn{1}{|c|}{} & \hspace{1ex} (messages); & \hspace{1ex} node; & - source of\\
\multicolumn{1}{|c|}{} &  & - not fault-tolerant & \hspace{1ex} randomness;\\
\hline
\multicolumn{1}{|c|}{} & - computation of & - less accurate; & \\
\multicolumn{1}{|c|}{\bf Digests} & \hspace{1ex} complex aggregates; & - resources needed & - local knowledge\\
\multicolumn{1}{|c|}{} & - result at all nodes; & \hspace{1ex} (e.g. larger & \hspace{1ex} of neighbor IDs;\\
\multicolumn{1}{|c|}{} &  & \hspace{1ex} messages); & \\
\hline
\end{tabular}
\caption{Summary of the characteristics of main data aggregation classes.}
\label{tab:main_agg_class_prop}
\end{table}

One should notice that most of the existing approaches only allow the computation of simple aggregation functions, such as \textsc{average}, \textsc{sum} and \textsc{count}, or others that can be derived from their combination (by executing multiple instances of the used algorithm). In many cases, this kind of aggregation functions is enough, but in many other situations the computation of more complex aggregation functions is more useful. A simple example can be found considering some load balancing application that aims to distribute equitably the global load of a system. In this case, the knowledge of the total or average load does not provide enough information to assess the distribution of the system load, i.e. determine if some processing nodes are overloaded or idle. Even the computation of the maximum and minimum is insufficient, although it allows the detection of a gap between the global load distribution, as it does not provide information about the number of processes at each load level. In this situation, an estimation the statistical load distribution is required to provide the desired information and reveal outlier values. Other examples can be found in the context of monitoring applications. For instance, in WSN estimating the distribution of the monitored attribute can be very useful to distinguish isolated sensor anomalies from the occurrence of a relevant event characterized by a certain amount of abnormal values. 
Few approaches are available to compute more complex aggregates and able to approximate the statical distribution of some attribute (see \ref{ssub:digests}). Existing algorithms from this class (i.e. digests) are more resource consuming and less accurate than other approach, so that their application should be carefully evaluated despite their additional value.


\section{Final Remarks and Future Directions} 
\label{sec:final_remarks_and_future_directions}

This survey was organized around three main contributions. First, it provides a formal definition of the target aggregation problem, defining different type of aggregations function and their main properties. Second, a taxonomy of the of the existing algorithms is proposed, from two perspectives: communication and computation, and the most relevant algorithms are succinctly described. Finally, a summary of the characteristics of the main approaches is provided, giving some guidelines about their suitability to different scenarios. 

Distributed data aggregation has been an active field of research in the last decade, and a huge diverse amount of techniques can be found in the literature. For this reasons, this survey intends to be an important time saving instrument, for those that desire to get a quick and comprehensive overview of the state of the art on distributed data aggregation. Moreover, by carefully highlighting the strength and limitations of the more pertinent approaches, this study can provide a useful assistance to help readers choose which technique to apply in specific settings. 

Currently, there is no ideal general solution to the distributed computation of an aggregation function, all existing techniques have its pitfalls (some more than others). Therefore, more research in this field will be expected in the next few years. In particular, due to the added value of computing complex aggregates, new algorithms might arise to estimate the statistical distribution of values, as the few existing approaches exhibit some limitations in terms of accuracy and resource consumption. Additional research efforts should be made to improve the support to churn, message loss, and continuous estimation of mutable input values.



\bibliographystyle{plain}
\bibliography{survey_data_agg_pj}


\end{document}